\documentclass[prb,aps,twocolumn,showpacs,amsmath,amssymb,citeautoscript]{revtex4}

\usepackage{graphicx}
\usepackage{amsmath}
\usepackage{times}

\newcommand{\rmS}{{\rm S}}
\newcommand{\rmR}{{\rm R}}
\begin{document}

\title{Ground State and Excitations of Quantum Dots with ``Magnetic Impurities''}

\author{Ribhu K. Kaul,$^{1,2}$ Denis Ullmo,$^{3}$ 
Gergely Zar\'and,$^{4,5}$ Shailesh Chandrasekharan,$^{1}$ and 
Harold U. Baranger$^{1}$}  
\affiliation{$^1$Department of Physics, Duke University, Box 90305, Durham, NC 27708, USA}
\affiliation{$^2$Institut~f\"ur~Theorie~der~Kondensierten~Materie,~Universit\"at~Karlsruhe,~76128~Karlsruhe,~Germany} 
\affiliation{$^3$CNRS and Universit\'e Paris-Sud, LPTMS, 91405 Orsay, France}
\affiliation{$^4$Research~Institute~of~Physics,~Technical~University~Budapest,~Budapest,~H-1521,~Hungary,} 
\affiliation{$^5$Institut~f\"ur~Theoretische~~Fesk\"orperphysik,~Universit\"at~Karlsruhe,~76128~Karlsruhe,~Germany}

\date{June 11, 2009; published as Phys. Rev. B. \textbf{80}, 035318 (2009)}

\begin{abstract}
We consider an ``impurity'' with a spin degree of freedom coupled to a finite reservoir of non-interacting electrons, a system which may be realized by either a true impurity in a metallic nano-particle or a small quantum dot coupled to a large one. We show how the physics of such a spin impurity is revealed in the many-body spectrum of the entire finite-size system; in particular, the evolution of the spectrum with the strength of the impurity-reservoir coupling reflects the fundamental many-body correlations present. Explicit calculation in the strong and weak coupling limits shows that the spectrum and its evolution are sensitive to the nature of the impurity and the parity of electrons in the reservoir. The effect of the finite size spectrum on two experimental observables is considered. First, we propose an experimental setup in which the spectrum may be conveniently measured using tunneling spectroscopy. A rate equation calculation of the differential conductance suggests how the many-body spectral features may be observed. Second, the finite-temperature magnetic susceptibility is presented, both the impurity susceptibility and the local susceptibility. Extensive quantum Monte-Carlo calculations show that the local susceptibility deviates from its bulk scaling form. Nevertheless, for special assumptions about the reservoir -- the ``clean Kondo box'' model -- we demonstrate that finite-size scaling is recovered. Explicit numerical evaluations of these scaling functions are given, both for even and odd parity and for the canonical and grand-canonical ensembles.
\end{abstract}


\maketitle


\section{Introduction}

The Kondo problem describes a single magnetic impurity interacting with a sea of electrons \cite{HewsonBook}. At temperatures $T$ of the order of or less than a characteristic scale, $T_{\rm K}$, the dynamics of the impurity and the sea of electrons become inextricably entangled, thus making Kondo physics one of the simplest realizations of a strongly correlated quantum system. In its original context, the impurity was typically an element of the 3\textit{d} or 4\textit{f} series of the periodic table, embedded in the bulk of a metal such as \textit{Cu} with \textit{s} conduction electrons. With the subsequent development of fabrication and control of micro- and nano-structures, it was pointed out \cite{Glazman88,Ng88} that a small quantum dot with an odd number of electrons -- small enough that its mean level spacing $\Delta_\rmS$ is much larger than the temperature -- could be placed in a regime such that it behaves as a magnetic impurity.\cite{GlazmanHouches05,ZarandRev06,GoldhaberGRev07} The first experimental implementations of this idea were naturally made by connecting the ``magnetic impurity'' formed in this way to macroscopic leads \cite{Goldhaber98,Cronenwett98,Pustilnik01,GoldhaberGRev07}. The flexibility provided by the patterning of two dimensional electron gas makes it possible, however, to design more exotic systems, by connecting the small magnetic impurity dot to larger dots playing the role of the electron reservoirs. Schemes to observe, for instance, two-channel SU(2) \cite{Oreg03,Potok06} or SU(4)\cite{Borda03,LeHur03,LeHur04,Galpin05,LeHur07,Choi05,Makarovski07a,Makarovski07b} Kondo have been implemented.

When the bulk electron reservoir of the original Kondo problem is replaced by a finite reservoir, two energy scales are introduced: the Thouless energy $E_{\rm   Th}$ associated with the inverse of the time of flight across the structure, and the mean level spacing $\Delta_\rmR$ \cite{Kouwenhoven97,AkkerMontamBook,Argaman93}.  A natural question which arises is therefore how these two new scales affect the Kondo physics under investigation.

Because a quantum impurity problem has point-like interactions, the local density of states $\rho_{\rm loc}(\epsilon) \!=\! \sum_\alpha |\phi_\alpha(0)|^2\delta(\epsilon-\epsilon_\alpha)$ completely characterizes the non-interacting sea of electrons ($\epsilon_\alpha$ and $\phi_\alpha$ are the one-body eigenvalues and eigenfunctions of the reservoir). For $T,T_{\rm K} \!\gg\! E_{\rm Th},\Delta_\rmR$, thermal smearing washes out the effects of both mesoscopic fluctuations and the discreteness of the reservoir spectrum. Indeed, in this regime, one may safely approximate $\rho_{\rm loc}$ by a constant $\rho_0$; the impurity behaves in much the same way as if it were in an infinite reservoir. In contrast, when $T, T_{\rm K} \lesssim E_{\rm Th},\Delta_\rmR$, the impurity senses the finiteness of the reservoir through the structure of $\rho_{\rm loc}(\epsilon)$. The presence of these new energy scales (which are ubiquitous \cite{Kouwenhoven97,AkkerMontamBook} in reservoirs made from quantum dots) is hence an essential and interesting part of Kondo physics in nano-systems and deserves to be understood thoroughly.

The implications of a finite Thouless energy, and of the associated mesoscopic fluctuations taking place in the energy range $[\Delta, E_{\rm Th}]$, have been investigated mainly in the high temperature range $T \!\gg\! T_{\rm K}$, where a perturbative renormalization group approach is applicable \cite{Zarand96,Kettemann04,KaulEPL05,Yoo05,UllmoRPP08} (see also related work \cite{Kettemann06,Kettemann07,Zhuravlev08} in the context of weakly disordered system). Less is known about the implications of mesoscopic fluctuations in the temperature range $T \!<\! T_{\rm K}$.

There is on the other hand already a much larger body of work concerning the ``clean Kondo box'' problem \cite{Thimm99}, namely the situation where mesoscopic fluctuations are ignored (or absent as may be the case in some one dimensional models), and only the existence of a finite mean level spacing is taken into account.  Simon and Affleck \cite{Simon02} and Cornaglia and Balseiro \cite{Cornaglia03} have, for instance, considered how transport properties are modified if one dimensional wires of finite length are inserted between the macroscopic leads and  the quantum impurity.  Ring geometries \cite{Affleck01,Simon01,Lewenkopf05,Simon06}, including the configuration corresponding to a two channel Kondo effect \cite{Simon06}, have also been investigated.

The basic Kondo box configuration, namely a quantum impurity connected to an electron reservoir with a finite mean level spacing, turns out to be  already a non-trivial problem and so has been investigated by various numerically intensive techniques such as the non-crossing approximation \cite{Thimm99} or the numerical renormalization group \cite{Cornaglia02a}.  In Refs.\,\onlinecite{KaulPRL06} and\,\onlinecite{Simon06} it was pointed out, however, that as only the regime $T \ll \Delta_R$ is affected by the finiteness of $\Delta$, a lot of physical insight  could be obtained by the analysis of the low energy {\em many-body} spectrum  of the Kondo box system (i.e.\ the ground state and first few excited states). An analysis of this low energy many-body spectra and of an experimental setup in which it could be probed was given in Ref.\,\onlinecite{KaulPRL06}.

\begin{figure}[t]
\includegraphics[width=2.4in]{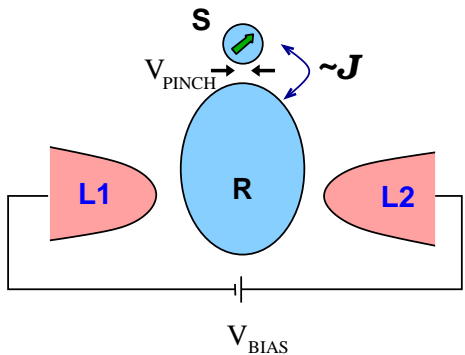}
\caption{(Color online) A double dot system, coupled very weakly to leads (${\rm L1, L2}$). In the Coulomb blockade regime, the small dot ${\rm S}$ behaves like a magnetic impurity that is coupled to a finite reservoir ${\rm R}$ provided by the large dot. The leads are used to measure the excitation spectrum of the system.}
\label{fig:setup}
\end{figure}

In this article, we would like on the one hand to provide a more detailed account of some of the analysis sketched in Ref.\,\onlinecite{KaulPRL06}, and furthermore to present an additional physical application, namely the low temperature magnetic response of the Kondo box system. See also Ref.\,\onlinecite{Pereira08} for an analysis of the addition energy of a Kondo box.

Since our focus is the consequences of a finite $\Delta_\rmR$, we consider the simplest possible configuration: a double dot system with a small dot acting as the magnetic impurity and a larger one playing the role of the electron reservoir, as illustrated in Fig.\,\ref{fig:setup}. The Hamiltonian describing this double-dot system is 
\begin{equation}
\label{eq:basic_ham}
H_{\rm R\text{-}S} = \sum_{\alpha \sigma} \epsilon_\alpha 
c^{\dagger}_{\alpha \sigma}c_{\alpha\sigma}
+ E_C( N^\rmR - {n^\rmR_g})^2
+ H^{\rm K,A}_{\rm int} \;.
\end{equation}
Here $c^\dagger_{\alpha \sigma}$ creates a state $\phi_{\alpha}(r)$ with spin $\sigma$ and energy $\epsilon_\alpha$ which is an exact one-body energy level in the bigger quantum dot R. These states include all the effects of static disorder and boundary scattering. $N^\rmR$ is the number operator for electrons in the reservoir, $n^\rmR_g$ the dimensionless gate voltage applied to the large dot R, and $E_C$ its charging energy. As the charging energy is the leading interaction for electrons in a finite system, we shall neglect all other interactions among the electrons on R. (See, e.g., Ref.\,\onlinecite{Murthy05,RotterPRL08,Pereira08} for work that includes interactions among electrons in R.) The last term in Eq.\,(\ref{eq:basic_ham}) contains the description of the small dot and the interaction between the dots.

We consider two models for the magnetic impurity quantum dot and its interaction with the reservoir R. For most of this paper, we use a ``Kondo''-like model, which therefore includes charge fluctuations only implicitly. In this case, the smaller quantum dot is represented by a spin operator $\mathcal{S}$. The interaction with the screening reservoir R is given by the usual Kondo interaction,
\begin{equation}
\label{eq:basic_ham_kondo}
H^{\rm K}_{\rm int} = J \;{\mathbf {\mathcal S}}\cdot
{\mathbf s} (0) \; ,
\end{equation}
describing the anti-ferromagnetic exchange interaction between the two dots, with $ {\bf s}(0) \!=\! \frac 1 2 f^{\dagger}_{0\sigma} {\overrightarrow{\sigma}_{\sigma\sigma^\prime}} f_{0\sigma}$ the spin density in the large dot at the tunneling position ${\bf r} \!\equiv\! 0$ and $f^{\dagger}_{0\sigma}\!\equiv\! \sum_{\alpha} \phi_{\alpha}(0) c^\dagger_{\alpha}$.

We also consider (see Sec.~\ref{sec:gndstatethm}) a multi-orbital ``Anderson''-type model that explicitly includes the effect of charge fluctuations on the quantum dot S:
\begin{eqnarray}
\label{eq:basic_ham_and}
H^{\rm A}_{\rm int} &=& \sum_{m \sigma} \epsilon^d_m d^{\dagger}_{m
 \sigma}d_{m\sigma}+ \sum_{m\sigma}t_{m}[f_{0\sigma}^\dagger
d_{m\sigma}\nonumber + {\rm H.c.} ] \\ &+&  
 U (N^{\rm S} - n^{\rm S}_g)^2.
\end{eqnarray}
Here the quantum dot S is described by a set of spin-degenerate energy levels $\epsilon^d_m$ created by $d^\dagger_{m \sigma}$ which couple to the state $f^\dagger_{0\sigma} \!\equiv\! \sum_\alpha \phi_\alpha(0) c^\dagger_{\alpha\sigma}$ in R. Interactions are included through the usual charging term of strength $U$, where $N^{\rm S}\!\equiv\! \sum_{m\sigma}d^\dagger_{m\sigma}d_{m\sigma}$ and $n^\rmS_g$ is the dimensionless gate voltage applied to the small dot. When $m$ takes only a single value, this reduces to the usual single-level Anderson model. The crucial feature of this model is that the R-S tunneling term (proportional to $t$) involves \textit{only one} state in the reservoir.

For temperature $T$ much larger than not only the mean level spacing $\Delta_\rmR$ but also the corresponding Thouless energy of the reservoir dot, the discreetness of the spectrum as well as mesoscopic fluctuations in R can be ignored. Thus one expects to recover the traditional behavior of a spin-1/2 Kondo or Anderson model. If $T \!\ll\! \Delta_\rmR$, however, significantly different behavior is expected. A simplifying feature of this limit is that many physical quantities can be derived simply from properties of the ground state and low-lying excited states.

To study the low temperature regime, we shall therefore in a first stage consider the low energy (many-body) spectrum of the Hamiltonian Eq.\,(\ref{eq:basic_ham}). Specifically, in Sec.~\ref{sec:gndstatethm} we extend (slightly) a theorem from Mattis \cite{Mattis67} that enables us to infer the ground state spin of the system. Using weak and strong coupling perturbation theory, we then construct in Sec.~\ref{sec:finitesize} the finite size spectrum of the Kondo problem in a box.

In a second stage, we consider a few observable quantities that are derived simply from the low energy spectra. We start in Sec.~\ref{sec:rate_eq} with tunneling spectroscopy, obtained by weakly connecting two leads to the \textit{reservoir} dot (Fig.\,\ref{fig:setup}). Using a rate equation approach, we predict generic features in the non-linear $I$-$V$ of our proposed device. We then address in Sec.~\ref{sec:susceptibilities} the low temperature magnetic response of the double dot system, and in particular discuss the difference between local and impurity susceptibilities which, although essentially identical for $T \!\gg\! \Delta_\rmR$, differ drastically when $T \!\ll\! \Delta_\rmR$. A further issue that we study is that the charging energy in R fixes the number of electrons rather than the chemical potential; thus, the canonical ensemble must be used rather than the grand-canonical. Use of the canonical ensemble accentuates some features in the susceptibility. Finally, we conclude in Sec.~\ref{sec:conclusions}.


\section{Ground State Theorem}
\label{sec:gndstatethm}

We now prove an exact ground state theorem for the models defined in Eqs.\,(\ref{eq:basic_ham})-(\ref{eq:basic_ham_and}): the ground state spin of the system is fixed, and in particular cannot depend on the coupling between the small dot and the reservoir. We give the value of this ground state spin in a variety of cases. 

The theorem is mainly an extension of a theorem due to Mattis \cite{Mattis67}. It relies on the fact that in a specially chosen many-body basis, all the off-diagonal matrix elements of these Hamiltonians are non-positive. It is then possible to invoke a theorem due to Marshall \cite{Marshall55,AuerbachBook} to infer the ground state spin -- a proof of Marshall's sign theorem is in Appendix~\ref{appendix:marthm}.

\subsection{``Kondo''-type models}

\begin{figure}[t]
\includegraphics[width=2.2in]{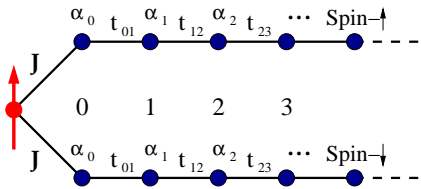}
\caption{(Color online) Mapping of the Kondo problem into a spin-chain with an impurity. The site '0' is the point in the bath that interacts with the impurity, and is used as the site to begin the tri-diagonalization of the one-body Hamiltonian of the bath. The sites are labeled by the integer $i$, in the sequence they are generated by the tri-diagonalization procedure. After tri-diagonalization, we are left with non-interacting electrons that feel an on-site potential $\alpha_i$ and can hop {\em only} to the neighboring sites with amplitude $t_{i,i+1}$.
} 
\label{fig:wilson}
\end{figure}

The starting point of the proof is the tri-diagonalization of the
one-body Hamiltonian of the reservoir, $H_\rmR$. Beginning with the
state $f_{0\sigma}$, one rewrites $H_\rmR$ as a one dimensional chain
with only nearest-neighbor hopping \cite{WilsonRMP75,HewsonBook}. This
transformation is illustrated in Fig.\,\ref{fig:wilson}. In this
one-body basis, the ``Kondo''-type model
Eqs.\,(\ref{eq:basic_ham})-(\ref{eq:basic_ham_kondo}) can be rewritten
as a sum of a diagonal and off-diagonal part:
\begin{equation}
\label{eq:dod}
H_{\rm R\text{-}S} = H_{\rm D} + H_{\rm OD} \;,
\end{equation}
\begin{equation}
  H_{\rm D} =  J {\mathcal S^z} s^z (0) + \sum_{i\sigma}
  \alpha_i \;f_{i\sigma}^{\dagger } f_{i\sigma}+ E_C( N^R-{n^\rmR_g})^2 \;,
\end{equation}
\begin{eqnarray}
\label{eq:offdiag} 
  H_{\rm OD} &=& -\frac{|{J}|}{2}\bigl[{\mathcal S^+} s^-(0)+ {\rm H.c.} \bigr]
\\ 
&&- \sum_{i\sigma} \big(|t_{i, i+1}|\;
f_{i\sigma}^{\dagger}f_{i+1\sigma} + {\rm H.c.}\big) \;. \nonumber 
\end{eqnarray}

Condition $1$ of the the Marshall theorem  
requires us to find a many-body basis in which all the off-diagonal matrix elements are non-positive. Consider the following basis:
\begin{equation}
\label{eq:basis}
| \Psi_\alpha \rangle = (-1)^{m-\mathcal{S}}
f_{i_{N^{\rm R}_\uparrow}{\uparrow}}^{\dagger}
... 
f_{i_1\uparrow}^{\dagger} 
f_{j_1\downarrow}^{\dagger}
... 
f_{j_{N^{\rm R}_\downarrow}{\downarrow}}^{\dagger} 
| 0 \rangle \otimes | m \rangle
\end{equation}
with $m$ the quantum number of ${\mathcal S}^z$ of the local spin and the site labels (positive integers) ordered according to $i_1<\!\dots\!<i_{N^{\rm R}_{\uparrow}}$ and $j_1<\!\dots\!<j_{N^{\rm R}_{\downarrow}}$. Note that this basis is diagonal with respect to both the total number of electrons in R, $N^{\rm R} \!=\! N^{\rm R}_\uparrow \!+\! N^{\rm R}_\downarrow$, and the $z$-component of the total magnetization, $S^z_{\rm tot} \!=\! m \!+\! (N^{\rm R}_\uparrow \!-\! N^{\rm R}_\downarrow)/2$.

The off-diagonal matrix elements come from two terms, the spin-flip term and the fermion hopping. With regard to the fermion hopping term, first, since the fermions have been written as a one-dimensional chain, there is no sign from the fermionic commutation relation. Additionally, one can use the freedom to choose the phase that defines the one-body states $f^\dagger_{i\sigma}$ to make the hopping integrals $t_{i,i+1}$ negative. Since the number of phases is the same as the number of hopping integrals $t_{i,i+1}$, all the $t_{i,i+1}$ can be made negative, as in Eq.\,(\ref{eq:offdiag}). This ensures that all off-diagonal matrix elements of the fermion hopping term in the many-body basis, Eq.\,(\ref{eq:basis}), are non-positive. With regard to the spin-flip term, note that its sign in $H_{\rm OD}$ can be fixed by rotating the spin $\mathcal{S}$ by an angle $\pi$ about the $z$-axis. In order to ensure that the off-diagonal elements due to the spin-flip term are negative, we have to include the additional phase factor $(-1)^{m-\mathcal{S}}$ appearing in the definition of the basis states in Eq.\,(\ref{eq:basis}).

Since the basis Eq.\,(\ref{eq:basis}) is diagonal in $N_\rmR$ and $S_{\rm tot}^z$, we will work in a fixed $(N_\rmR,S_{\rm tot}^z)$ sector. Condition (ii) of Marshall's theorem -- 
connectivity of the basis states by repeated application of $H_{\rm R\text{-}S}$ -- is easily seen to be satisfied for the ``Kondo'' model for all ${J}\!\neq\! 0$, in a given $(N_\rmR,S_{\rm tot}^z)$ sector. However, when $J\!=\!0$, condition 2 is violated: the impurity spin cannot flip and hence some basis states in a $(N_\rmR,S_{\rm tot}^z)$ sector cannot be connected to each other by repeated applications of $H_{\rm R\text{-}S}$.

\begin{table}
\begin{tabular}{|c|c|c|c|}\hline
  $\mathcal{S}$& ${J}$ &$N_\rmR$&Spin of $|G\rangle$
\\\hline
 1/2& ANTI&ODD&0\\\hline
 1/2& ANTI&EVEN&1/2\\\hline
 1&ANTI&EVEN& 1\\\hline
 1&ANTI& ODD& 1/2\\\hline
 1/2& FERRO&ODD&1\\\hline
 1/2& FERRO&EVEN&1/2\\\hline
\end{tabular}
\caption{Ground state spins for different Kondo problems according to the theorem combined with perturbation theory. Marshall's theorem adapted to the model defined by Eqs.\,(\ref{eq:basic_ham}) and (\ref{eq:basic_ham_kondo}) says that the ground state spin does not change in a parametric evolution of the Hamiltonian. The only exception is the crossing of the point $J\!=\!0$, hence the sign of $J$ appears in the table. 
}
\label{fig:spintable}
\end{table}

We have thus shown that the Kondo model satisfies the two conditions of Marshall's theorem in a given $(N_\rmR,S_{\rm tot}^z)$ sector. Now note that given $N_\rmR$ and $\mathcal{S}$, the competing spin multiplets for the ground state spin ($S_{\rm tot}$) can either be integer spin multiplets or half-integer spin multiplets. Suppose for instance they are integer multiplets (this is true, e.g., when $N_\rmR$ is odd and $\mathcal{S}\!=\!1/2$). Marshall's theorem guarantees that in the $S_{\rm tot}^z\!=\!0$ sector the lowest eigenvalue can never have a degeneracy; this ensures that in a parametric evolution there can never be a crossing in the $S_{\rm tot}^z\!=\!0$ sector. Since each competing multiplet has a representative state in the $S_{\rm tot}^z\!=\!0$ sector, we infer that the ground state spin does not change as the coupling $J$ is tuned. This is true as long as we do not cross the point $J\!=\!0$, because this point (as explained above) violates condition 2 in the proof of the theorem. Hence the ground state spin can be different for ferromagnetic and anti-ferromagnetic $J$, but does not change with the magnitude of the coupling: the ground state spin for all $J$ may hence be inferred by lowest order perturbation theory in $J$. The ground state spin for a few representative cases is displayed in Table~\ref{fig:spintable}.

\subsection{``Anderson''-type models}

We can prove a similar theorem for the model defined by Eqs.\ (\ref{eq:basic_ham}) and~(\ref{eq:basic_ham_and}). We begin by tri-diagonalizing the electrons in the reservoir R, as for the Kondo case. In addition we have to tri-diagonalize the electrons in the quantum dot S, a process which begins with the state 
$\tilde d_{0\sigma} \!=\! (1/t_{\rm R\text{-}S})\sum_{m}t_{m} d^\dagger_{m\sigma}$ 
where $t_{\rm R\text{-}S}\!=\!(\sum_m t_m^2)^{1/2} $. An organization of 
$H_{\rm R\text{-}S}$ into diagonal and off-diagonal parts then yields
\begin{eqnarray}
\label{eq:dod_and}
  H_{\rm D} &=& \sum_{m, \sigma} \alpha^d_m \tilde d^{\dagger}_{m \sigma}\tilde d_{m\sigma} +  U (N^{\rm S} - n^{\rm S}_g)^2\\
&&+ \sum_{i,\sigma} \alpha_i \;f_{i\sigma}^{\dagger } f_{i\sigma}+ E_C( N^{\rm R}-{n^{\rm R}_g})^2  \nonumber\\ 
  H_{\rm OD} &=& - |t_{\rm R\text{-}S}|(f_{0\sigma}^\dagger \tilde d_{0\sigma}\nonumber + {\rm H.c.} )
\nonumber \\ 
&&- \sum_{m,\sigma}(|t^d_{m, m+1}|\; \tilde d_{m\sigma}^{\dagger}\tilde d_{m+1\sigma} + {\rm H.c.})\nonumber \\
&&- \sum_{i,\sigma}(|t_{i, i+1}|\; f_{i\sigma}^{\dagger}f_{i+1\sigma} + {\rm H.c.}) \;.
 \end{eqnarray}
We note again that the sign of all the hopping integrals can be fixed as displayed above by an appropriate selection of the arbitrary phase that enters the definition of the $\tilde d^\dagger_{m \sigma}$ and the $f^\dagger_{i\sigma}$. The appropriate basis that has only non-positive off-diagonal matrix elements is, then, simply
\begin{eqnarray}
\label{eq:basis_and}
| \Psi_\alpha \rangle &=&
f_{i_{N^\rmR_\uparrow}{\uparrow}}^{\dagger}
... 
f_{i_1\uparrow}^{\dagger}
\tilde d_{k_{N^{\rm S}_\uparrow}{\uparrow}}^{\dagger}
... 
\tilde d_{k_1\uparrow}^{\dagger}\nonumber\\
& & \tilde d_{l_1\downarrow}^{\dagger}
... 
\tilde d_{l_{N^{\rm S}_\downarrow}{\downarrow}}^{\dagger} 
f_{j_1\downarrow}^{\dagger}
... 
f_{j_{N^\rmR_\downarrow}{\downarrow}}^{\dagger} 
| 0 \rangle \;.
\end{eqnarray}
The total number of particles is now 
$N^{\rm tot}\!=\! N^\rmR_\uparrow+N^\rmR_\downarrow+N^{\rm S}_\uparrow+N^{\rm S}_\downarrow$, and the $z$-component of spin is 
$S^z_{\rm tot}\!=\!(N^\rmR_\uparrow+N^{\rm S}_\uparrow - N^\rmR_\downarrow - N^{\rm S}_\downarrow)/2$.

In the case of the Anderson-type model Eqs.\,(\ref{eq:basic_ham}) and (\ref{eq:basic_ham_and}), the result for the ground state spin is remarkably simple: the ground state spin has $S^{\rm tot}\!=\!0$ for $N^{\rm tot}$ even and $S^{\rm tot}\!=\!1/2$ for $N^{\rm tot}$ odd. There is no possibility of having a ground state spin other than the lowest.


\section{Finite Size Spectrum}
\label{sec:finitesize}

In this section, we outline the main features of the low-energy finite-size spectrum for the Kondo problem, Eqs.\ (\ref{eq:basic_ham})-(\ref{eq:basic_ham_kondo}). The basic idea is to use perturbation theory around its two fixed points: at the weak coupling fixed point ($J\!=\!0$) expand in $J$, and at the strong coupling fixed point expand in the leading irrelevant operators (Nozi\`eres' Fermi-liquid theory). We begin by analyzing the classic case of $\mathcal{S}\!=\!1/2$ with anti-ferromagnetic coupling, and then turn to the under-screened Kondo problem realized by anti-ferromagnetic coupling and $\mathcal{S}\!=\!1$.

\subsection{$\mathcal{S}=1/2$: Screened Kondo problem }
\label{sec:1/2Kondo}

\begin{figure}
\includegraphics[width=2.0in]{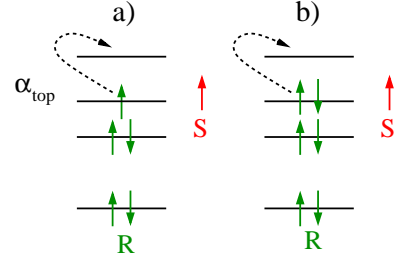}
\caption{Weak-coupling perturbation theory: schematic illustration of the unperturbed system for (a)~$N^\rmR$ odd and (b)~$N^\rmR$ even. The spin marked ${\mathcal S}$ is that of the small quantum dot while the solid lines represent the spectrum of the (finite) reservoir. The dashed lines show the lowest energy orbital excitation in each case; note that in the even case, any excitation requires promoting an electron to the next level and so involves a minimum energy of order $\Delta_\rmR$.
}
\label{fig:weakcplg}
\end{figure}

\textit{Weak-Coupling Regime:} In the weak coupling regime defined by $\Delta_\rmR \!\gg\! T_{\rm K}$, given a realization of the reservoir R, we can always make $J$ small enough so that the spectrum can be constructed through lowest order perturbation theory.

The unperturbed system for $N$ odd is shown schematically in Fig.\,\ref{fig:weakcplg}(a). At weak coupling the eigenstates follow from using degenerate perturbation theory in all the multiplets of the unperturbed system. The ground state and the first excited state are obtained by considering the coupling 
\begin{equation} \label{eq:topmost}
  H_{\rm top} = J |\phi_{\alpha_{\rm top}}(0)|^2 \; {\rm \bf s}_{\rm top} 
  \cdot {\mathcal S}
\end{equation}
where ${\rm \bf s}_{\rm top}$ is the spin of the topmost (singly occupied)
level $\alpha_{\rm top}$ of the large dot. The ground state is
therefore a singlet ($J\!>\!0$) and the first excited state is a triplet with excitation energy
\begin{equation} \label{eq:EST}
\delta E_{\rm ST} = J |\phi_{\alpha_{\rm top}}(0)|^2 \ll
\Delta_\rmR \; .
\end{equation}
The next excited states are obtained by creating an electron-hole excitation in the reservoir [shown as a dashed arrow in Fig.\,\ref{fig:weakcplg}(a)]. Combining the spin $1/2$ of the reservoir with that of the small dot, one obtains a singlet of energy $\sim\! \Delta_\rmR$ separated from a triplet by a splitting $\sim\! \mathcal{J} \Delta_\rmR/4$, where we define $\mathcal{J} \!=\! J\rho$ with $\rho = \langle |\phi_\alpha(0)|^2 \rangle/\Delta_\rmR$ the mean local density of states.

In the $N$ even case depicted in Fig.\,\ref{fig:weakcplg}(b), the ground state is trivially a doublet. The first excited eigenstate of the unperturbed system is an $8$-fold degenerate multiplet obtained by promoting one of the bath electrons to the lowest available empty state [shown as a dashed arrow in Fig.\,\ref{fig:weakcplg}(b)]. As $J$ is turned on, this multiplet gets split into two $S_{\rm tot} \!=\! 1/2$ doublets and one $S_{\rm tot} \!=\! 3/2$ quadruplet. In general the two doublets have lower (though unequal) energy than the quadruplet.

\begin{figure}
  \includegraphics[width=2.0in]{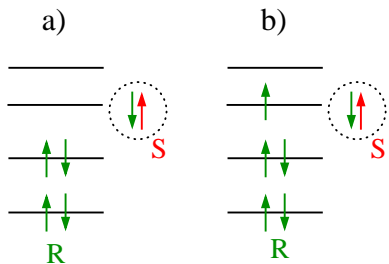}
  \caption{Strong-coupling perturbation theory: schematic illustration of the unperturbed system for (a)~$N^\rmR$ odd and (b)~$N^\rmR$ even. For $\mathcal{S}=1/2$, a conduction electron is bound to the impurity at strong coupling. This leaves effectively a gas of even (odd) weakly interacting quasi-particles. We note here that in the usual case of $T_{\rm K}\ll D$ (bandwidth of the reservoir), the formation of a singlet between the impurity spin and the reservoir for $T_{\rm K} \!\gg\! \Delta_\rmR$ is a complicated many-body effect: it is {\em not} just a singlet between the spin and the topmost singly occupied level, as is the case for very weak coupling.
}
\label{fig:strngcplg}
\end{figure}

\textit{Strong-Coupling Regime:} For $T_{\rm K} \!\gg\! \Delta_\rmR$,
on the other hand, the impurity spin $\mathcal S$ is screened by the
conduction electrons, and we can use Nozi\`eres' ``Fermi-liquid''
theory.\cite{Nozieres74,Nozieres78} In the very strong coupling limit, one
electron is pulled out of the Fermi sea to bind with the impurity;
this picture essentially holds throughout the strong coupling-regime
\cite{GlazmanHouches05,Nozieres74,Nozieres78}.
For $N$ odd (even) one
ends up effectively with an {\em even (odd) number} of quasi-particles
that interact with each other only at the impurity site through a
repulsive effective interaction
\begin{equation}
   U_{\rm FL} \sim (\Delta_\rmR^2/T_{\rm K}) \, 
      n_{\uparrow}(0) \,n_{\downarrow}(0)
\label{eq:U_FL}
\end{equation}
which is weak ($T_{\rm K} \!\gg\! \Delta_{\rm R}$).
The quasi-particles have the same mean level spacing $\Delta_\rmR$ as the original electrons, but the spacing between two quasi-particle levels is not simply related to the spacing of the original levels in the chaotic quantum dot. This case is illustrated in Fig.\,\ref{fig:strngcplg}.

For $N$ odd, the ground state is thus a singlet (as expected from our theorem), and the excitations start at energy $\sim\! \Delta_\rmR$ since a quasi-particle must be excited in the reservoir. The first two excitations consist of a spin $S_{\rm tot}\!=\!1$ and a $S_{\rm tot}\!=\!0$. Because the residual quasi-particle interaction is repulsive, the orbital antisymmetry of the triplet state produces a lower energy; the splitting is about $\sim\! \Delta_\rmR^2/T_{\rm K}$.

In the $N$ even case at strong coupling, there are an odd number of quasi-particles in the reservoir, and so the ground state is a doublet. The first excited multiplet must involve a quasi-particle-hole excitation in the reservoir. There are two such excitations that involve promotion by one mean level spacing on average (either promoting the electron in the top level up one, or promoting an electron in the second level to the top level). Thus, the first two excitations are doublets.

\begin{figure}
\includegraphics[width=3.0in]{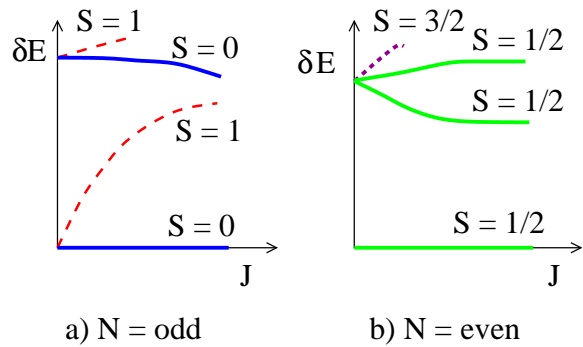}
\caption{(Color online) Schematic of the energy eigenvalues of the double-dot system as a function of the coupling $J$ for (a)~$N$ odd and (b)~$N$ even in the $\mathcal{S}\!=\!1/2$ anti-ferromagnetic coupling case. Energy differences are shown with respect to the ground state which therefore appears on the $x$-axes. The relation to the double-dot experiment Fig.\,\ref{fig:setup} is that the $y$-axis here is like $V_{\rm BIAS}$ and the $x$-axis is like $V_{\rm PINCH}$. The excitations will show up as peaks in the differential conductance $G$.
}
\label{fig:illust}
\end{figure}

\textit{Crossover between Weak- and Strong-Coupling:} Remarkably, the ordering of the $S_{\rm tot}$ quantum numbers of the ground state and two lowest excitations is the same in both the $T_{\rm K} \!\gg\! \Delta_\rmR$ and $T_{\rm K} \!\ll\! \Delta_\rmR$ limits. It is therefore natural to assume that the order and quantum numbers are independent of $T_{\rm K}/\Delta_\rmR$. Thus we arrive at the schematic illustration in Fig.\,\ref{fig:illust}.

\subsection{$\mathcal{S}=1$: Under-screened Kondo problem }

The theorem and perturbation theory analysis presented above has an interesting generalization to the under-screened Kondo effect, in which $\mathcal{S}\!>\!1/2$. We will consider for concreteness the case $\mathcal{S}\!=\!1$. Note that the under-screened Kondo effect has been realized experimentally in quantum dots.\cite{VanDerWiel02}

\begin{figure}
\includegraphics[width=3.0in,clip]{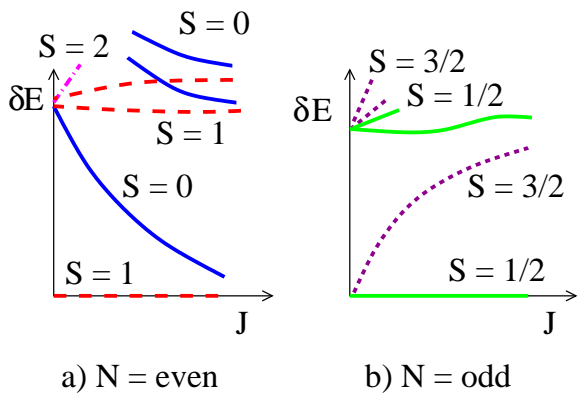}
\caption{(Color online) Schematic illustration of the finite-size spectrum of the under-screened Kondo problem, $\mathcal{S}\!=\!1$ with anti-ferromagnetic coupling, for (a) $N$ even and (b) $N$ odd. In each panel we show all the excitations up to order $\Delta$, both at strong and weak coupling. When it seems plausible, we have connected the strong and weak coupling limits; note the necessity of crossings among the excited states in (a).
}
\label{fig:illust_us}
\end{figure}

For $N$ even and ${\mathcal S}\!=\!1$, we find that the ground state for all $J$ has $S_{\rm tot}\!=\!1$. At weak coupling, this follows directly from perturbation theory -- the reservoir has spin zero and $J$ is too small to promote an electron so the spin of the ground state is just that of the small dot. The theorem then implies that ${\mathcal S}\!=\!1$ for all $J$. The first excited multiplet is at energy of order $\Delta_\rmR$. It splits into a singlet, two triplets, and a quintuplet; as $J$ increases, the singlet has the lowest energy because the coupling is anti-ferromagnetic. 

In the opposite limit of strong coupling, as $J\!\rightarrow\!\infty$, one electron from R binds to the impurity spin forming a spin-1/2 object. For $J\!=\!\infty$, this spin does not interact with the quasi-particles in R; however, when $J\!\neq\! \infty$, the flow to strong coupling generates other irrelevant operators that connect the spin to the quasi-particles. It is known from studies of the under-screened Kondo problem that the leading irrelevant operator is a \textit{ferromagnetic} Kondo coupling \cite{Nozieres80} (the sign of the coupling follows heuristically from perturbation theory in $t/J$). However, since one of the electrons is bound to the spin, there is an \textit{odd} number of quasi-particles in the effective low energy ferromagnetic Kondo description -- the level filling is as in Fig. \ref{fig:strngcplg}(b). Since the ferromagnetic Kondo problem flows naturally to \textit{weak} coupling \cite{HewsonBook}, we are again justified in doing perturbation theory in the coupling, and so recover that the ground state has $S_{\rm tot}\!=\!1$. From the small ferromagnetic coupling, we conclude that the first excited state is a singlet separated from the ground state by an asymptotically small energy (as $T_{\rm K}/\Delta_\rmR \!\rightarrow\! \infty$). The next excited state involves promotion of a quasi-particle to the next level within R and so has energy of order $\Delta_\rmR$. It is a triplet because of the ferromagnetic coupling, with a nearby singlet in the strong coupling limit. Note that there are two possible quasi-particle excitations with energy of order $\Delta_\rmR$ (as discussed in the $\mathcal{S}\!=\!1/2$ case), and so two singlet-triplet pairs.

The proposed crossover from weak to strong coupling for $N$ even is shown in Fig.\,\ref{fig:illust_us}(a). Note that in this case, level crossings of excited states \textit{must} occur: the two singlets at energy of order $\Delta_\rmR$ at strong coupling come from energies greater than $\Delta_\rmR$ at weak coupling, and so cross the $S \!=\! 2$ state. The two singlet-triplet pairs at strong coupling are shown to be slightly different because each involves a different level spacing; thus, there is an additional level crossing as one of the singlets comes below a triplet.

In the $N$ odd case, weak anti-ferromagnetic coupling implies that the ground state spin is $S_{\rm tot}\!=\!1/2$. The first excited state is the other multiplet involving no excitations in the reservoir, $S_{\rm tot}\!=\!3/2$. The next excited states are the $S_{\rm tot}\!=\!1/2$ and $3/2$ states that involve promoting one electron by one level. In the strong coupling limit, we repeat the mapping to a ferromagnetically coupled impurity, yielding this time an even number of quasi-particles in the reservoir. Now the first excited state involves promoting a quasi-particle in the reservoir by one level; the ferromagnetic coupling implies that the $S_{\rm tot}\!=\!3/2$ state has the lowest energy among the possible multiplets. Making again the reasonable assumption that the two limits are connected to each other in the simplest manner possible, we arrive at the schematic illustration in Fig.\,\ref{fig:illust_us}(b). In contrast to the $N$ even case, no level crossings are definitely required.

We stress here that the evolution of the finite size spectra shown in Figs.\,\ref{fig:illust} and \ref{fig:illust_us} are totally different in each of the cases illustrated. The finite-size spectrum is hence an interesting way to observe the Kondo effect in nano-systems, each impurity problem having its own unique spectrum. 


\section{Non-linear I-V characteristics of the R-S system}
\label{sec:rate_eq}

We now turn to the question of how to observe the features of the finite size spectrum delineated in the previous section. Any physical observable depends, of course, on the spectrum of the system and so could be used as a probe. We choose to concentrate on two: (1) In the next section, we discuss the \textit{magnetic susceptibility} of the R-S system, a classic quantity in Kondo physics. (2) In this section we discuss the \textit{conductance} across the device shown in Fig.\,\ref{fig:setup}. The advantage of this physical quantity is that the finite-size spectrum can be observed directly in the proposed experiment. The emphasis here is on transfer of electrons entirely by real transitions; cotunneling processes, which involve virtual states, are briefly discussed at the end of the section.  

A current through the R-S system clearly involves number fluctuations on it. For a general value of the gate voltage [$n^{\rm R}_g$ in Eq.\,(\ref{eq:basic_ham})], however, the ground state will have a fixed number of electrons, and hence $G\!=\!0$ (Coulomb blockade). When $V_{\rm BIAS}=V_{\rm 1} - V_{\rm 2}$ is increased sufficiently, the Coulomb blockade is lifted, and $G(V_{\rm BIAS})$ has a sequence of peaks. It is possible to extract the excitation spectrum of the R-S system from the position of these peaks\cite{VonDelft01}. In principle, there is a peak in $G$ for every transition $\alpha \rightarrow \beta$ that involves a change in $N$. 
As discussed in subsection~\ref{sec:C} we shall however choose a particular setting such that only a limited number of these transitions play a role, making in this way simpler the reconstruction of the underlying low-energy many-body spectra.

\subsection{Method} 

In order to describe transport through the R-S system (realized through either
a double dot or a metallic grain with a single magnetic impurity), we solve
the appropriate rate equations for the real transitions
\cite{Beenakker91,VonDelft01}. The rate equations are a limit of the quantum
master equation in which the off-diagonal elements of the density matrix are
neglected. The dynamics of the quantum dot can then be described simply by the
probability $P_{\alpha}$ that the R-S system is in a given many-body state
$\alpha$. In thermal equilibrium these $P_{\alpha}$ are the
Boltzmann weights. The electrons in the leads are assumed to always be in
thermal equilibrium; hence, the probability that a given one-body state in the
leads is occupied is given simply by the Fermi-Dirac function $f(\epsilon)
\!\equiv\! 1/(e^{\epsilon/T}\!+\!1)$. Here, $\epsilon$ is the deviation from
the electro-chemical potential $E_{\rm F}\!+\!V_{1,2}$, where $V_1$ and $V_2$
are the voltages on leads $1$ and $2$ respectively. 


Steady state requires that the $P_{\alpha}$ are independent of time. Hence, the various rates of transition from $\alpha$ to $\beta$, $\Lambda_{ \beta \alpha}$, must balance, leading to a linear system for the $P_{\alpha}$, 
\begin{equation}
\label{eq:ratebal}
\sum_{\beta} \Lambda_{\alpha \beta} P_{\beta} = \sum_{\beta}
\Lambda_{\beta \alpha} P_{\alpha} \;. 
\end{equation}
In addition, the occupation probabilities should be normalized, $\sum_{\alpha} P_{\alpha} \!=\! 1$.

There are four transitions that have to be taken into account: addition or removal of an electron from lead L1 or L2. We denote the rates for these four processes as $\Lambda^{\pm {\rm L1, L2}}_{\alpha \beta}$, and the $\Lambda_{\alpha \beta}$ in Eq.\,(\ref{eq:ratebal}) are sums of these four transition rates. Once we have the $P_{\alpha}$ from (\ref{eq:ratebal}), the current is simply 
\begin{equation}
I_{\rm 2} \equiv \frac{dN_{\rm 2}}{dt} =\sum_{\beta, \alpha}
(\Lambda^{+\rm L2}_{\beta \alpha}- \Lambda^{-\rm L2}_{\beta \alpha})
P_{\alpha} \;. 
\end{equation}
The conductance $G$ then follows by differentiating $I(V_{\rm BIAS})$.

The rates $\Lambda_{\alpha \beta}$ can be calculated in second order perturbation theory in the reservoir-lead coupling term, using Fermi's golden rule \cite{VonDelft01}. For example, consider the addition of an electron to R-S from ${\rm L}_{\rm 1}$ corresponding to a transition $\beta (N) \!\rightarrow\! \alpha (N+1)$ on R-S: 
\begin{eqnarray}
\Lambda^{+\rm L1}_{\alpha \beta} 
&=& \frac{2 \pi \mathcal{V}_{\rm 1}^2}{\hbar}\int\!
d\epsilon\,\rho(\epsilon)\, f(\epsilon)\,
\delta(E_\alpha-E_\beta-\epsilon-V_{\rm 1})\nonumber\\ 
&=& \Gamma_{\rm 1} f(E_\alpha - E_\beta -V_{\rm 1}) \label{eq:FGR}
\end{eqnarray}
where $\Gamma_{\rm 1,2} = 2\pi \mathcal{V}_{\rm 1,2}^2\rho(E_{\rm F})/\hbar$. $\mathcal{V}_{\rm 1}$ is the amplitude for the above process. Although, in general it will have some dependence on $\alpha$ and $\beta$ as well as the coupling $\mathcal{J}$, we will ignore such dependence here. We will, however, retain the distinction between $\mathcal{V}_{\rm 1,2}$ and allow these to be tuned by the gates that define the R-L1 and R-L2 junctions. 

To summarize our approach, to find the conductance in the proposed tunneling experiment, we have solved the rate equations for transferring an electron from lead 1 to the reservoir and then to lead 2.\cite{Beenakker91,VonDelft01} We assume that (1) the coupling of the lead to each state in R is the same (mesoscopic fluctuations are neglected), (2) the Kondo correlations that develop in R-S do not affect the matrix element for coupling to the  leads,\cite{fnspacing} (3) there is a transition rate $\lambda_{\rm rel}$ that provides direct thermal relaxation between the eigenstates of R-S with fixed $N$, (4) the electrons in the lead are in thermal equilibrium, and (5) the temperature $T$ is larger than the widths $\Gamma_1,\Gamma_2$ of the R-S eigenstates due to L1 and L2.

\subsection{Magnetic Field}

A Zeeman magnetic field ${\bf B}_{\rm Z}$ can be used as an effective probe of the various degeneracies of the R-S system. We shall assume that the magnetic field does not couple to the orbital motion of the electrons:
\begin{equation} 
H_{\rm Z} = -g \mu_B {\bf B}_{\rm Z} \cdot {\bf  S} \;.
\label{eq:ZeemanHam}
\end{equation}
This can be achieved in the semiconductor systems by applying the field parallel to the plane of motion of the electrons. The effect of an orbital magnetic field in ultra-small metallic grains is argued to be small in Ref.\,\onlinecite{VonDelft01} for moderate fields.  

We may neglect the effect of $B_{\rm Z}$ on the lead electrons: The only characteristic of the lead electrons appearing in the rate equation calculations is the density of states at $E_{\rm F}$. All that $B_{\rm Z}$ does to the lead electrons is to make the modification $\rho(E_{\rm F}) \!\rightarrow\! \rho(E_{\rm F}\pm g\mu_B B_{\rm Z}/2)$. Since the band is flat and wide (on the scale of $B_{\rm Z}$) to an excellent approximation, this has no effect. 

The effect of $B_{\rm Z}$ on the R-S system is complicated if the $g$-factors for the ${\rm S}$ and ${\rm R}$ electrons are different, as would be the case for a magnetic impurity in a metallic nano-particle. If we assume, however, that the $g$-factors for the electrons on the ${\rm S}$ and ${\rm R}$ systems are the same, as is relevant for the semiconductor quantum dot case illustrated in Fig.\,\ref{fig:setup}, then $H_{\rm Z}$ becomes simply $-g \mu_B B_{\rm Z} S^z_{\rm tot}$. The energy of a given many-body level $\alpha$ is then $E_{\alpha} \!-\! g\mu_B B_{\rm Z} S^z_{\rm tot}$ where $S^z_{\rm tot}$ is the corresponding eigenvalue of the many-body state.

\subsection{Application to R-S System} 
\label{sec:C}

To identify characteristic features in the transport properties, let us analyze a situation in which only a limited number of transitions show up.\cite{VonDelft01} For a  $\mathcal{S} \!=\! 1/2$ Kondo problem the most interesting features appear in the spectrum when there is an {\em odd} number of electrons in the reservoir. These states appear clearly when an electron is added to a $N$ even reservoir and the parameters are such that the excited states of the $N\!+\!1$ electron reservoir dominate. 

We thus consider the following situation: For zero bias, assume that the R-S system is brought into a Coulomb blockade valley, not far from the $N\!\to\! N+1$ transition. This could be done by adjusting $V_1$ and $V_2$ in the setup of Fig.\,\ref{fig:setup} (with $V_1\!=\!V_2$), or more realistically with the the help of the additional gate voltage $n^\rmR_g$ in Eq.\,(\ref{eq:basic_ham}). We take this setup as the origin of the bias potentials ($V_1\!=\!V_2\!=\!0$). 
%
%
Upon applying a bias $V_1 \!>\! V_2$, electrons flow from lead 1 to lead 2.

We assume that the rates $\Gamma_{1,2}$ are sufficiently small that virtual processes (cotunneling) can be entirely neglected for this subsection; that is, all relevant transitions occur on shell and can be described by the Fermi golden rule expression Eq.\,(\ref{eq:FGR}). Furthermore, we take $\Gamma_2 \!\gg\! \Gamma_1$. Because $V_1 \!>\! V_2$, this means that it takes much longer to add an electron to the dot than to empty it. Thus, the dot tends to be occupied by $N$ electrons.

Several conditions are needed in order to restrict the discussion to just the lowest lying states of the system. First, we shall assume that $T \!\ll\! \Delta_R$ so that in equilibrium only the ground state $S\!=\!1/2$ doublet, with energy $E_G[N]$, needs to be considered. In a non-equilibrium situation, however, higher excited states $E_i[N]\!\sim\! \Delta_R$ can also be populated: the excess energy of the electron supplied by the bias can be used to leave the dot in an excited state. Apart from the ground state doublet, we take all excited states $E_i[N]$ to be higher in energy than $E_G[N+1]$.\cite{EFterm} Then, if an excited state is populated, it will quickly relax to an energy below $E_G[N+1]$ through a rapid exchange of particles back and forth between the dot and lead 2 ($\Gamma_2 \!\gg\! \Gamma_1$). Because off-shell processes are assumed negligible, this relaxation will stop as soon as an $N$-electron state $E_i[N]$ below $E_G[N+1]$ is reached. We  assume that the energy of the first excited $S\!=\!1/2$ doublet, $E_1[N]$, is large enough that $ E_1[N] + V_2 \!\!>  E_G[N+1] \!>\! E_G[N] + V_2 $. Then only the $N$-electron ground state multiplet needs to be retained in the calculation.

With regard to the $N\!+\!1$ electron states, we limit ourselves to a small enough bias such that only transitions to the three lowest excited multiplets need to be taken into account. In this way, only a small number of transitions  will show up in the excitation spectrum, making it relatively simple to analyze.\cite{VonDelft01}   

When a magnetic field is applied, note the following unusual behavior: Since there is no way to decay from the $S_z\!=\! -1/2$  state to the $S_z\!=\!+1/2$ state of the lowest doublet without involving virtual processes explicitly neglected here, the  doublet will remain out of equilibrium: the $S_z \!=\! -1/2$ state can be significantly populated even though  $g \mu_B B_{\rm Z} \gg k_B T$. 

\begin{figure}[t]
\includegraphics[width=2.8in,clip]{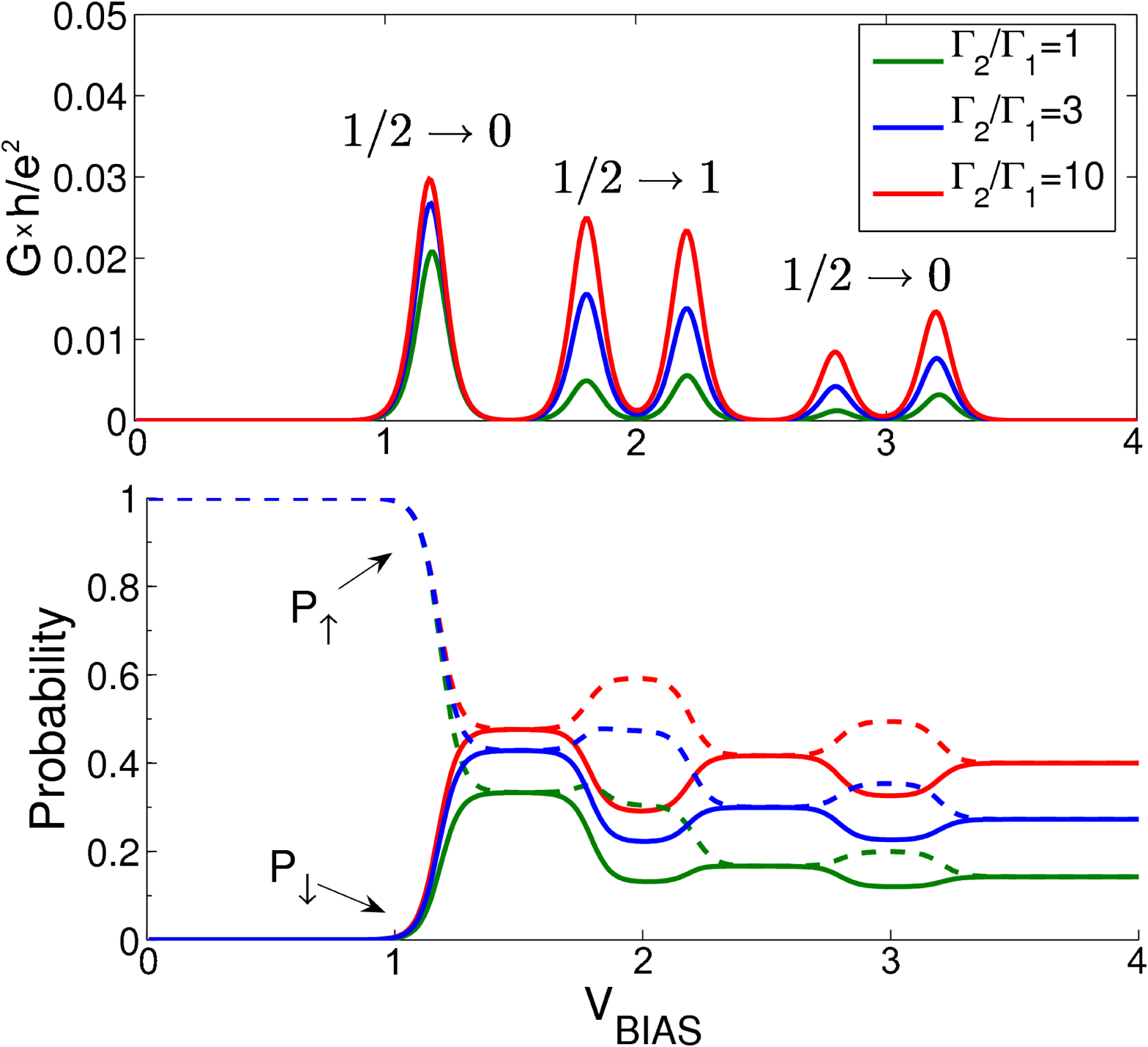}
\caption{ (Color online) Illustrative calculation of transport spectroscopy starting from the ground state
  of a $N$ even Kondo R-S system. Transitions are $S\!=\!1/2 \rightarrow
  S\!=\!0,1$ as marked. Top panel: The differential conductance as a function
  of bias voltage for different values of the asymmetry between $L,R$
  tunneling rates. Bottom panel: The probability of occupation of the two
  states forming the $S\!=\!1/2$ ground state. For large asymmetries
  $P_{\uparrow}+P_{\downarrow}\approx 1$, as expected. Even though $k_{\rm B} T/g \mu_B B_{\rm Z} \!=\! 0.2 \ll 1$, $P_{\downarrow}$ is large because this
  calculation neglects inelastic relaxation on the R-S system, allowing it
  to stay well out of equilibrium. $\Gamma_1 \!=\! 0.01$, $g\mu_B B_{\rm Z} \!=\! 0.4$, and $\Gamma_2/\Gamma_1 \!=\! 10$, $3$, and $1$ from top to bottom.\cite{fnGamma2Gamma1} 
}
\label{fig:ass1} 
\end{figure}

\begin{figure}[t]
\includegraphics[width=2.8in,clip]{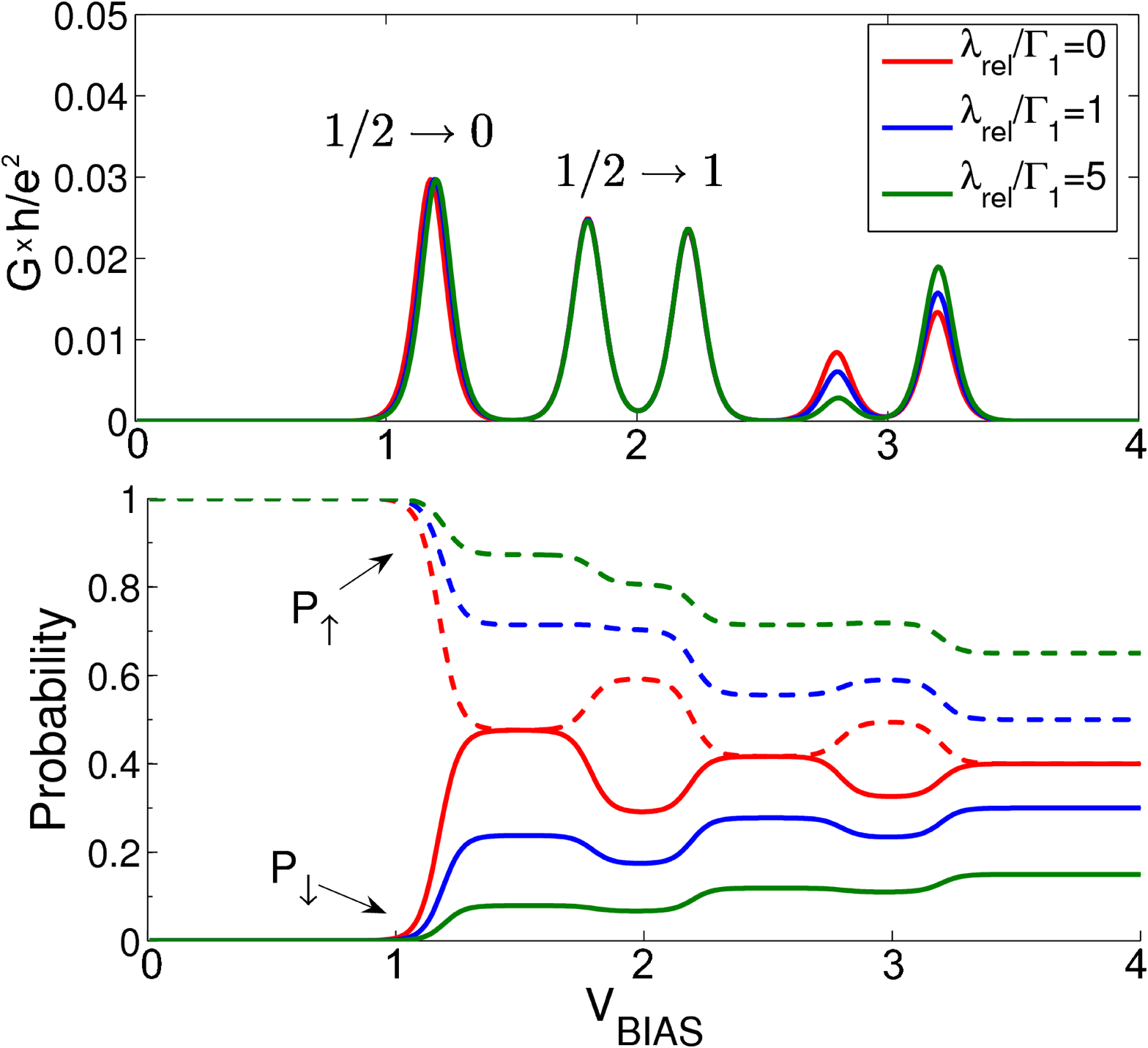}
\caption{(Color online) Transport spectroscopy with energy relaxation
  included. $\Gamma_2/\Gamma_1\!=\!10$ (fixed), the parameter $\lambda_{\rm
    rel}$ that models energy relaxation varies, and other parameters are as in
  Fig.\,\ref{fig:ass1}. Top panel: The differential conductance as a function
  of bias voltage. Note how the symmetry of the $S\!=\!1/2 \rightarrow
  S\!=\!1$ peaks is unaffected while the $S\!=\!1/2, S_z\!=\!-1/2 \rightarrow
  S\!=\!0$ transition is suppressed (this transition would completely vanish
  if the R-S system were in thermal equilibrium). Bottom panel: The
  probability of occupation of the two states forming the $S\!=\!1/2$ ground
  state. As $\lambda_{\rm rel}$ is increased, the R-S system has a larger
  probability to occupy its ground state ($S_z\!=\!1/2$). 
}  
\label{fig:ass2} 
\end{figure}

\subsection{Results for $G$} 

With the above assumptions, results for the differential conductance are shown in Fig.\,\ref{fig:ass1}. We assumed that the system parameters (gate potential, $V_1$, $V_2$, and $\delta E_{\rm ST}$) are such that the first three states of the $N\!+\!1$ electron system coincide with the ground state of the $N$ electron system for $V_{\rm BIAS} \!=\! 1$, $2$, and $3$, respectively, at $B\!=\!0$. (See Fig 4.\ of Ref.\,\onlinecite{KaulPRL06} for $dI/dV$ in the $B\!=\!0$ case.) We are thus assuming that the excited triplet state lies midway between the two lowest singlet states (see Fig.\,\ref{fig:illust}), placing ourselves in the middle of the cross-over regime.

First, note that the ground state to ground state transition, $S\!=\!1/2 \rightarrow S\!=\!0$, yields only one peak even at non-zero $B_{\rm Z}$. This is because the $S_z\!=\!-1/2$ state of the doublet cannot be populated before some current is flowing through the R-S system ($k_{\rm B} T/g \mu_B B_{\rm Z} \!=\! 0.2 \!\ll\! 1$). However, after the first transition, the $N+1$ state can decay into the $S_z=-1/2$ state. Hence, we expect the higher $S\!=\!1/2 \rightarrow S\!=\!0$ transitions to split in a magnetic field, as in Fig.\,\ref{fig:ass1}. 

The next feature to understand is the two $S\!=\!1/2 \rightarrow S\!=\!1$ transitions. These two peaks occur because out of the six transitions between the multiplets, two are forbidden by spin conservation and the other four split into two degenerate sets.

How is one then to distinguish between a $S\!=\!0$ and $S\!=\!1$ state, since they both split into two as a function of $B_{\rm Z}$? One possible method is to observe the peak heights in $G(V_{\rm BIAS})$ keeping $B_{\rm Z}$ fixed. These are plotted in Fig.\,\ref{fig:ass1} for a variety of $\Gamma_2/\Gamma_1$.\cite{fnGamma2Gamma1} A clear feature is that the two $S\!=\!1/2 \rightarrow S\!=\!0$ peaks are very asymmetric, while the $S\!=\!1/2 \rightarrow S\!=\!1$ are almost symmetric. This is for a robust physical reason: each $S\!=\!1$ peak gets contributions from both $S_z\!=\!1/2$ and $-1/2$ initial states, while in the $S\!=\!0$ transition each peak gets a contribution from only one, the $S_z\!=\!1/2$ for the taller peak and $S_z\!=\!-1/2$ for the shorter one. The associated probabilities, $P_{\uparrow}$ and $P_{\downarrow}$, are shown in the lower panel of Fig.\,\ref{fig:ass1}. Thus the peak heights in the $S\!=\!1$ transitions are insensitive to the difference between the probability of occupation of the two states in the doublet, while the peak heights in the $S\!=\!0$ transitions are sensitive to this difference.

\subsection{Energy Relaxation} 
In any real system, there are mechanisms of energy relaxation beyond the energy conserving exchange of electrons with the leads that is given by the rate equations. These mechanisms can involve, for instance, interactions with phonons or, more simply, higher order virtual processes between the R-S system and the 
leads that are neglected in the Fermi's golden rule approach Eq.\,(\ref{eq:FGR}). These relaxation processes are particularly important for the second $S\!=\!1/2 \rightarrow S\!=\!0$ transition. If the system is in perfect thermal equilibrium this transition should yield a single peak, even in the presence of a $B_{\rm Z} \neq 0$. The second peak is suppressed even if only on-shell processes are taken into account, as discussed above, but explicit energy relaxation causes this suppression to be more pronounced. 

To model energy relaxation, we include a transition rate between the $S_z \!=\! \pm 1/2$ states that satisfies detailed balance (i.e., with Boltzmann weights),
\begin{equation}
\Lambda_{\alpha \beta} = \lambda_{\rm rel} \frac{e^{-\varepsilon_\alpha/T}
}{e^{-\varepsilon_\alpha/T}+e^{-\varepsilon_\beta/T}}
\end{equation}
where $\varepsilon_{\alpha}$ is the energy eigenvalue of the $\alpha^{\rm th}$ state. 

The effect of this term is shown in Fig.\,\ref{fig:ass2}. Clearly as the relaxation rate is increased, the peak in the second $S\!=\!1/2\rightarrow S\!=\!0$ transition coming from non-equilibrium effects is suppressed further. Note, however, that the heights of the $S\!=\!1/2\rightarrow S\!=\!1$ transition are unaffected (in both relative and absolute magnitude).

\subsection{Cotunneling Spectroscopy}

While the approach proposed above should be reasonably simple to implement, because the excitations of both the $N$ and $N+1$ electron systems may come into play, the resulting experimental conductance curves may in some circumstances be non-trivial to interpret. Therefore, we mention, without going into detail, an alternative way to extract the excitation spectra from the differential conductance.  Though within the ``Coulomb blockade diamond'', on-shell processes such as the ones considered above are
forbidden by energy conservation constraints, a small current can nevertheless be measured, which is associated with virtual (cotunneling) processes.\cite{Aleiner02}

At very low bias, these virtual processes are necessarily elastic as the electron transferred from one lead to the other does not have enough energy to leave the R-S system in an excited state.  However, each time $V_{\rm BIAS}$ reaches a value corresponding to an excitation energy of the system with $N$ electrons, a new ``inelastic'' channel is open, as the electron has the option to leave the R-S system in an excited state as it leaves the structure.  The opening of these new channels produce steps in the differential conductance within the Coulomb diamond.  These steps are small, but clearly observable experimentally \cite{DeFranceschiPRL01,ZumbuhlPRL04,Makarovski06}. 

Because of the smallness of the associated currents, observing this substructure within the Coulomb diamond is certainly more challenging experimentally than for observing the main peaks associated with on-shell processes.  On the other hand, the time elapsed between the successive transfers of an electron across the structure is large enough that the initial state of the $N$-particle system is always the ground state.  If they can be measured accurately, the cotunneling steps within the Coulomb diamond may therefore lead more directly to the $N$-particle excitation spectra.

Summarizing this section, we have shown in detail how $dI/dV$ measurements enable one to extract the finite size spectrum and spin quantum numbers of the R-S system, using the case when the ground state has an even number of electrons as an example. In particular, we argued that the relative peak height of the Zeeman split terms ($B_{\rm Z} \!\neq\! 0$) reflects the spin quantum number of the excitation: asymmetric peak heights correspond to $S\!=\!0$, whereas symmetric peak heights correspond to $S\!=\!1$. The case when $N$ is odd is straightforward to analyze in a similar way. Transitions from $S\!=\!0$ to $S\!=\!1/2$ or $3/2$ can easily be distinguished: the former splits into two in a magnetic field while the latter splits into four.


\section{Magnetic response of the double~dot~system}
\label{sec:susceptibilities}

We turn now to studying a second physical observable which probes the finite size spectrum of the system, namely the magnetic susceptibility defined by
\begin{equation} 
\chi = \frac{1}{\beta} \frac{\partial^2 \log Z}{\partial B^2} \; ,
\end{equation}
where $Z$ is the canonical or grand-canonical partition function depending on the ensemble considered. As in Section \ref{sec:rate_eq}, we assume that the magnetic field is in plane so that only the Zeeman coupling needs to be considered, Eq.\,(\ref{eq:ZeemanHam}). We furthermore distinguish between the local susceptibility $\chi_{\rm loc}$, corresponding to the case where $B$ couples only to the quantum impurity spin (${\bf S} \equiv {\bf \mathcal S}$), and the situation where $B$ couples to the total spin of the R-S system (${\bf S} \equiv {\bf S}_{\rm tot}$). In the latter case, the impurity susceptibility $\chi_{\rm imp}$ is defined as the difference $\chi_{\rm tot} \!-\! \chi_0$ between the total magnetic response and that of R in the absence of the impurity dot. 


For a wide ($D \!\gg\! T_{\rm K})$ and flat ($T \!\gg\! E^R_{\rm Th}, \Delta_\rmR$) band the local and impurity susceptibilities are essentially identical \cite{AndClog}. Indeed the effect of the magnetic field on the reservoir electrons is just to shift the energies of the spin up electrons with respect to the spin down by a fixed amount. If the spectrum is featureless, this only affects in practice the edge of the band, which in the limit $\mathcal{J} \!\equiv\! J \rho \!\to\! 0$ and $D \!\to\! \infty$ with fixed $T_{\rm K}$ will not affect the Kondo physics. More precisely, for small but finite $J \rho$ (and again for a wide flat band) the impurity susceptibility, being associated with the correlator of a constant of the system, can be written as
\begin{equation}
 T_{\rm K} \chi_{\rm imp} = f_\chi(T/T_{\rm K}) + \mathcal{O} (T_{\rm K}/D) \;
\end{equation}
where $f_\chi$ is a universal function of the ratio $(T/T_{\rm K})$. On the other hand, since the spin ${\bf \mathcal S}$ of the impurity is not conserved, a multiplicative renormalization factor $z_{\chi_{\rm loc}}$ needs to be introduced for the local susceptibility so that $T_{\rm K} \chi_{\rm loc} = z_{\chi_{\rm loc}} f_\chi(T/T_{\rm K}) $. 
For $z_{\chi_{\rm loc}}$ we use a form motivated by two loop renormalization, $1/z_{\chi_{\rm loc}} =1-\mathcal{J}+\alpha\mathcal{J}^2$ for $\mathcal{J} \!\ll\! 1$, with the coefficient of the quadratic term determined empirically, $\alpha=-0.4$. (For a discussion of $z_{\chi_{\rm loc}}$ in the context of two loop renormalization, see e.g.\ Ref.\,\onlinecite{Barzykin98}.)
We note here that in the universal regime $T_{\rm K}/D \rightarrow 0$, one also has $\mathcal{J} \rightarrow 0$ and hence $z_{\chi_{\rm loc}}\rightarrow 1$. In practical numerics, even though $T_{\rm K}/D$ is small enough that the $\mathcal{O}(T_{\rm K}/D)$ correction can be neglected, $\mathcal{J} \simeq 1/\ln(D/T_{\rm K})$ need not be as small; hence, it is necessary to include the prefactor correction $z_{\chi_{\rm loc}}$ to observe good scaling behavior.

In the regime $T \!\ll\! \Delta_\rmR$ that we consider here, however, the reservoir electron spectrum is not featureless near the Fermi energy, and the Zeeman splitting of the conduction electrons affects in practice the whole band, and not just the band edge. One therefore does not particularly expect any simple relation between the local and impurity susceptibilities. We now discuss the behavior of these quantities in this regime. We start with the canonical ensemble, for which the number of particles $N$, and therefore the parity of $N$, is fixed. We'll consider in a second stage the grand canonical ensemble and so neglect charging effects in the reservoir [$E_C \!=\!0$ in Eq.\,(\ref{eq:basic_ham})]; in this case the spin degeneracy induces finite fluctuations of the particle number even in the zero temperature limit.

\subsection{Canonical ensemble}

Since $S^z_{\rm tot}$ is a good quantum number, the impurity susceptibility in the canonical ensemble follows immediately from the information contained in Fig.\,\ref{fig:illust}, i.e.\ from the knowledge of the total spin and excitation energy of the first few many-body states. Neglecting all the levels with an excitation energy of order $\Delta_\rmR$ (because $T \!\ll\! \Delta_\rmR$), we simply get for $\chi_{\rm tot}$ a spin 1/2 Curie law for even $N$, and a spin 1 Curie law damped by $\exp[-\beta \, \delta E_{\rm ST}]$ for odd N. In this latter case, the magnetic response in the absence of the impurity is also a spin 1/2 Curie law; thus, for $\delta E_{\rm ST} \!\ll\! \Delta_\rmR$ one finds
\begin{equation} 
\chi_{\rm imp} =  \left\{
\begin{array}{ll}
{\displaystyle \beta \frac{(g \mu_B)^2}{4} \;,} & \mbox{$N$ even} \\[0.1in]
\beta (g \mu_B)^2 
\left[ 2 \exp(-\beta\, \delta E_{\rm ST}) - \frac{1}{4} \right], &
\mbox{$N$ odd} 
\end{array}
\right.  \label{eq:chi_imp}
\end{equation} 

The local susceptibility on the other hand involves ${\mathcal S}$ which is not a conserved quantity. Its computation therefore requires knowledge of the eigenstates, in addition to the eigenenergies and total spin quantum numbers contained in Fig.\,\ref{fig:illust}. We can follow the same approach used in Section~\ref{sec:finitesize} and analyze the two limiting regimes of coupling between the reservoir and the impurity quantum dot. We will then use a numerical Monte Carlo calculation in the intermediate regime and investigate how well it is described by a smooth interpolation between the two limiting regimes.

In the weak coupling regime, $T_{\rm K} \!\ll\! \Delta_\rmR$, we assume that even if some renormalization of the coupling constant $\mathcal J$ takes place, the eigenstates are the ones obtained from first-order perturbation theory in this parameter. For even $N$ at $T \!\ll\! \Delta_\rmR$, the impurity spin decouples from the (frozen) electron sea, and one obtains again a spin 1/2 Curie law. For odd $N$ at $T \!\ll\! \Delta_\rmR$, the system formed by the impurity spin and the singly occupied orbital decouples from the set of doubly occupied levels. The magnetic response is the same as for two spin 1/2 particles interacting through Eq.\,(\ref{eq:topmost}). We thus obtain
\begin{equation} 
\label{eq:chi_loc_low}
\chi_{\rm loc} =  \left\{
\begin{array}{ll}
{\displaystyle \beta \frac{(g \mu_B)^2}{4} \;,} & \mbox{$N$ even} \\[0.1in]
{\displaystyle
\frac{(g \mu_B)^2 e^{-\beta\, \delta E_{\rm ST}}}{1 + 3 e^{-\beta\, \delta E_{\rm ST}}}
  \left[\frac{e^{+\beta\, \delta E_{\rm ST}}-1} {2\,\delta\!E_{\rm ST}} +
             \frac{\beta}{2}  \right]
,} & \mbox{$N$ odd}
\end{array} \right.
\end{equation} 
valid for $T_{\rm K} \!\ll\! \Delta_\rmR$.

Turning now to the strong coupling regime, we follow Nozi\`eres' Fermi liquid
picture \cite{Nozieres74,Nozieres78}, where low energy states  $\vert \Psi \rangle$ (with $E_\Psi \!\ll\! T_{\rm K}$) are constructed from quasiparticles which interact locally according to Eq.\,(\ref{eq:U_FL}). In a local magnetic field, the energy of a state  $\vert \Psi \rangle$ is modified to 
\begin{equation}
E_\Psi(B) =  (g \mu_B B_{\rm Z}) \langle \Psi | {\mathcal S}_z | \Psi \rangle 
          - (g \mu_B B_{\rm Z})^2 \sum_{\xi \neq \psi}
          \frac{|\langle \xi | {\mathcal S}_z | \Psi \rangle
            |^2}{E_\xi - E_\Psi}
\label{eq:Epsi_strong}
\end{equation}
where the sum is over all the many-body excited states $\xi$. The first
term in this expression yields the effect of a change of the quasiparticle phase shift on the energy. It is important when $N$ is even: one of the quasiparticle states is singly occupied, and its energy is shifted by an amount $\sim\! (g \mu_B B_{\rm Z}/T_K) \Delta_R$. Thus the system acts like a spin-1/2 particle with an effective $g$-factor given by $g\Delta_R /T_K$. 
The result is a weak Curie susceptibility $\sim\! (g \mu_B \Delta_R/T_K)^2 / 4T$ at low temperature. 

The second term captures the effect of electron-hole quasiparticle excitations. It produces a non-zero contribution even when the discreteness of the spectrum is ignored, as may be seen as follows. First, the density of states of particle-hole excitations of energy $\Delta E$ in a Fermi liquid is proportional to $\Delta E$. In a Kondo state, the density of single-particle states is increased by a factor of $1/T_K$ because of the Kondo resonance. Thus we may replace the sum in Eq.\,(\ref{eq:Epsi_strong}) by an integral using a density of states proportional to $\sim\! \Delta E/T_K^2$. The integral should be cutoff at an energy of order $T_K$, where the Kondo resonance ends. Thus the second term in Eq.\,(\ref{eq:Epsi_strong}) gives a contribution $\sim\! (g \mu_B B_{\rm Z})^2/T_K $ to the energy, and a corresponding contribution $\sim\! (g \mu_B)^2/T_K$ to $\chi_{\rm loc}$.

Note that this second contribution is independent of the finite-system parameter $\Delta_R$ and so is the universal (bulk) part of the local susceptibility.\cite{FiniteSize_chiloc} It should behave smoothly as $T/\Delta_\rmR$ becomes smaller than one. In particular, if one considers a system without mesoscopic fluctuations (i.e.\ with constant spacings and wave function amplitudes at the impurity), we expect to recover the bulk behavior $T_{\rm K} \chi_{\rm loc} = z_{\chi_{\rm loc}} f_\chi(T/T_{\rm K}) $  for $\Delta_\rmR/T_K\to 0$. For $N$ even this relation holds only if the weak Curie behavior of the first term is not too large; more precisely, we expect $\chi_{\rm loc}$ to follow the universal behavior as long as
$T \agt \Delta_R^2/T_K$.

With these arguments, we have thus arrived at a complete description of the
magnetic susceptibility in both the weak and strong coupling limits for the
canonical ensemble. Note in particular the difference in the conditions for
having $\chi_{\rm loc} \!\to\! \chi_{\rm bulk}$ from those for having
$\chi_{\rm imp} \!\to\! \chi_{\rm bulk}$. Both limits hold in the regime $T
\!\gg\! \Delta_\rmR$ no matter what the value of $T_{\rm K}$. However in
addition, $\chi_{\rm loc} \!\to\! \chi_{\rm bulk}$ for any $T$ as long as
$\Delta_\rmR/T_K\to 0$.

\subsection{Grand-canonical ensemble}
\label{sec:grand-canonical}

Use of the grand canonical ensemble [which involves neglecting charging effects in the reservoir, $E_C \!=\!0$ in Eq.\,(\ref{eq:basic_ham})] introduces additional complications compared to the canonical ensemble case above.
To illustrate, recall first the behavior in the absence of the impurity, i.e.\ for a system of independent particles occupying doubly degenerate states $\epsilon_\alpha$. For $T \!\gg\! \Delta_\rmR$ in the grand canonical ensemble, the magnitude of the fluctuation of the number of particles will be significantly larger than one. Thus, even if the canonical ensemble result for $N$ even is quite different from that for $N$ odd [Eqs.\,(\ref{eq:chi_imp})-(\ref{eq:chi_loc_low})], such an odd-even effect would be completely washed out here: whatever the choice of the chemical potential $\mu$, configurations with odd or even $N$ would be as probable.

In the low temperature regime on the other hand, as soon as $T \!\ll\! \min_\alpha(|\mu-\epsilon_\alpha|)$ (which is usually $\sim\! \Delta_\rmR$), there is a fixed, even number of particles in the system. It is possible to make the average number of particles odd by choosing $\mu \!=\! \epsilon_{\alpha_F}$ for some orbital $\alpha_F$. In that case, as $T/\Delta_\rmR \!\to\! 0$, all orbitals $\alpha \!<\! \alpha_F$ are doubly occupied, all orbitals $\alpha \!>\! \alpha_F$ are empty, and independent of $T$ the orbital $\alpha_F$ has probability $1/4$ to be empty, $1/4$ to be doubly occupied, and $1/2$ to be singly occupied. For quantities showing some odd-even effect in the canonical ensemble but no strong dependence on $N$ once the parity is fixed (such as the local susceptibility), the grand canonical ensemble produces a behavior which is the average of the the odd and even canonical response, even though the mean number of particles $\langle N \rangle$ is odd.

Turning now to the full R-S system, the above non-interacting picture should certainly still hold in the weak coupling regime. If, either by adjusting $\mu$ or by making use of some symmetry of the one particle spectrum, $\langle N \rangle$ is kept fixed with an even integer value as $T \!\ll\! \Delta_\rmR$, one should recover the canonical magnetic response for even $N$. In contrast, the magnetic response for odd $\langle N \rangle$ should be the average of the canonical odd and even responses.

In the strong coupling regime (following again Nozi\`eres' Fermi liquid description), one also has an essentially non-interacting picture, but with effectively one less particle since one reservoir electron is used to form the Kondo singlet. The role of ``odd'' and ``even'' are then exactly reversed from the weak coupling case: for $T \!\ll\! \Delta_\rmR$ the grand canonical response will be the average of the canonical odd and even response for \textit{even} $\langle N \rangle$, and will be exactly the canonical response for \textit{odd} $\langle N \rangle$.

\subsection{Universality in a clean box}

The previous discussion mainly addressed the two limiting behaviors -- weak and strong coupling. To investigate the intermediate regime we now turn to numerical calculations. In particular, we use the efficient continuous-time quantum Monte Carlo algorithm introduced in Ref.\,\onlinecite{Yoo05}, with in addition adaptations to compute quantities in the canonical ensemble \cite{canon_method}. 
We study the behavior of the singlet-triplet gap $\delta E_{\rm ST}$ and the local susceptibility $\chi_{\rm loc}$; the impurity susceptibility $\chi_{\rm imp}$ follows directly from $\delta E_{\rm ST}$.

To focus on the consequences of the discreetness of the one particle spectrum while avoiding having to explore an excessively large parameter space, we  disregard the mesoscopic fluctuations of the spectrum and the wave-functions. That is, we consider the simplified ``clean Kondo box'' model \cite{Thimm99,Simon02,Cornaglia03} defined by $\phi_\alpha(0) \!\equiv\! \rho \Delta_\rmR$ and $\epsilon_{\alpha+1} \!-\! \epsilon_{\alpha} \!\equiv\! \Delta_\rmR$ independent of $\alpha$. For initial results for the more realistic ``mesoscopic Kondo model'' see Refs.\,\onlinecite{KaulEPL05} and\,\onlinecite{Yoo05}.

Under these conditions, the problem is described by only three dimensionless parameters: the coupling ${\mathcal J} \!=\! J\rho$, and the two energy ratios $D/\Delta_\rmR$ and $T/\Delta_\rmR$. For small ${\mathcal J}$ and large $D$ ($\gg\! \Delta_\rmR, T$), ${\mathcal J}$ and $D$ can be scaled away in the usual manner so that, except for renormalization prefactors such as $z_{\chi_{\rm loc}}$ which may still contain some explicit dependence on $\mathcal J$, physical quantities depend on $\mathcal J$ and $D$ only through the Kondo temperature $T_{\rm K}$.

We therefore expect, again up to the factor $z_{\chi_{\rm loc}}$, that both susceptibilities for the ``clean Kondo box'' model will be universal functions of the two parameters $T/\Delta_\rmR$ and $T_{\rm K}/\Delta_\rmR$. This function may, however, be different for the local and impurity susceptibilities, and will also depend on the parity and type of ensemble considered.

Before discussing how well the limiting behaviors discussed above describe the whole parameter range, we shall first check that our numerics confirm the expected universality. For the impurity susceptibility, since Eq.\,(\ref{eq:chi_imp}) is valid in the full range of coupling as long as $T \!\ll\! \Delta_\rmR$, we only need to verify that for $N$ odd the singlet-triplet excitation energy $\delta E_{\rm ST}$ is, as expected from the same argument, a universal function of $(T_{\rm K}/\Delta_\rmR)$: $\delta E_{\rm ST}/\Delta_\rmR \!=\! F(T_{\rm K}/\Delta_\rmR)$ with the limiting behaviors
\begin{equation}
   F(x) \approx \left\{
\begin{array}{ll}
    1,            & x \gg 1 \\
 - 1/{\rm ln}(x), & x \ll 1 \; .
\end{array}
\right. \label{eq:FST}
\end{equation}
The strong coupling behavior follows directly from the discussion in Sec.~\ref{sec:1/2Kondo}. The scaling behavior of $F_{\rm ST}$ at weak-coupling can, on the other hand, be obtained from a perturbative renormalization group argument: in the perturbative expression for $\delta E_{\rm ST}$ Eq.\,(\ref{eq:EST}), replace the coupling constant ${\mathcal J}$ by its renormalized value ${\mathcal J}_{\rm eff}$ at the scale $\Delta_\rmR$. Within the one-loop approximation, ${\mathcal J}_{\rm eff} \!=\! {\mathcal J} / [1 - {\mathcal J} \ln(D/\Delta_\rmR)]$ yields
\begin{equation}
\delta E_{\rm ST} \approx \frac{\mathcal{J} \Delta_\rmR}
{1-{\mathcal{J}}\ln(D/\Delta_\rmR)} \; .
\end{equation}
Substituting the one-loop expression for the Kondo temperature,
$T_{\rm K} \!=\! D\exp(-1/{\mathcal J})$, now yields the second line of (\ref{eq:FST}).

In the crossover regime, we find $F(x)$ through continuous time quantum Monte Carlo (QMC) calculations using a modification of the algorithm presented in Ref.\,\onlinecite{Yoo05} with updates which maintain the number of particles (canonical ensemble) \cite{canon_method}. To extract $\delta E_{\rm ST}$, we measure the fraction $P$ of states with $({S^z_{\rm tot}})^2\!=\!1$ visited in the Monte-Carlo sampling at temperature $T$. For a fixed ${\mathcal J}$ and large $\beta\!=\!1/T$, $P(\beta)$ can be excellently fit to the form ${2}/({ 3 + e^{\beta\, \delta E_{\rm ST}}})$ valid for a two-level singlet-triplet system. Repeating this procedure yields $\delta E_{\rm ST}$ for a variety of ${\mathcal J}$ and $\Delta_\rmR$.

\begin{figure}[t]
\centering
\includegraphics[width=3.37in,clip]{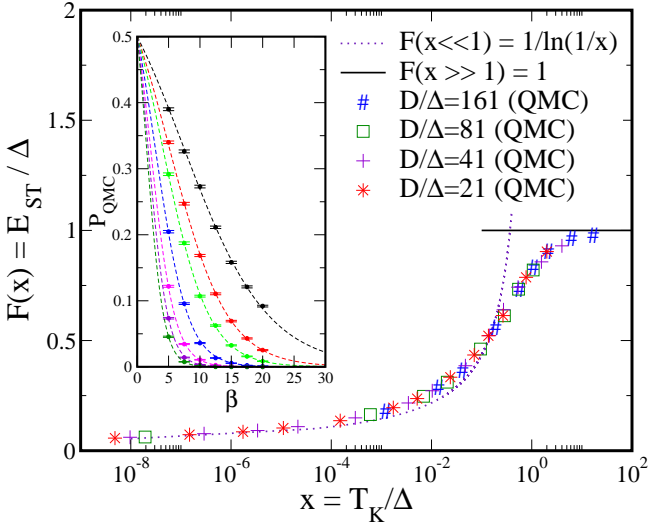}
\caption{(Color online) Universal form of the singlet-triplet gap ($N$ odd) in the absence of mesoscopic fluctuations (``Kondo in a box model''). Note the excellent data collapse for a wide range of bare parameters upon plotting the extracted gaps as a function of $T_{\rm K}/\Delta_\rmR$. Inset: $P_{\rm QMC}(\beta)$ for $D/\Delta_\rmR \!=\!41$. Circles are data for fixed ${\mathcal J}\!=\!0.05$, $ 0.10$, $ 0.13$, $ 0.15$, $ 0.19$, $ 0.23$, $ 0.26$, and $ 0.30$ (right to left). 
Lines are one parameter fits to the two state singlet-triplet form used to extract the values of the gap.}
\label{fig:EST}
\end{figure}

\begin{figure*}
\includegraphics[width=4.2in]{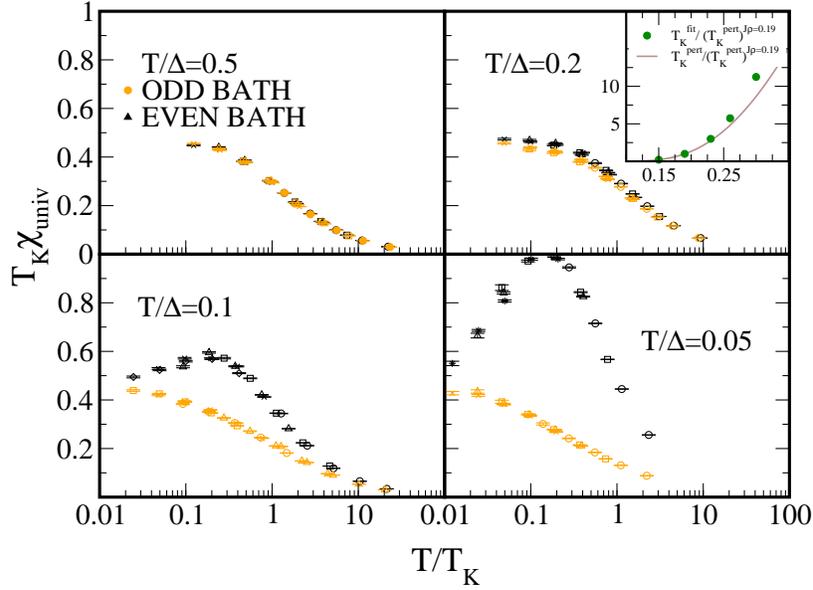}
\caption{(Color online) Rescaled susceptibility $\chi_{\rm univ} =
  z^{-1}_{\rm loc} \chi_{\rm loc} $ [in units of
$(g \mu_B)^2/4$] in the grand canonical ensemble as a function of
$T/T_{\rm K}$ while keeping the ratio $T/\Delta_\rmR$ fixed. The four
panels correspond to $T/\Delta_\rmR \!=\!0.5,\,0.2,\,0.1$, and
$0.05$. In each panel, the light (yellow online) and dark symbols
correspond respectively to odd and even mean number of particles in
the bath. Note the collapse of the data on universal curves, and the
substantial difference between the even odd cases at low
temperature. Inset: Comparison between the Kondo temperatures
extracted from our fits and the two-loop perturbative formula. As
expected, the agreement is excellent for $\mathcal{J} \!\ll\! 1$. At
larger values of $\mathcal{J}$, the two-loop formula tends to
underestimate $T_{\rm K}$.  
In each panel there is data for $D/\Delta_\rmR\!=\!20(21),\,40(41),\,80(81),\,160(161),\,320(321)$ for the odd (even) case, and for each case we have used $\mathcal{J}\!=\!0.19,\,0.23,\,0.26$, and $0.30$.
}
\label{fig:univsus_gc} 
\end{figure*}

\begin{figure*}
\includegraphics[width=6.0in]{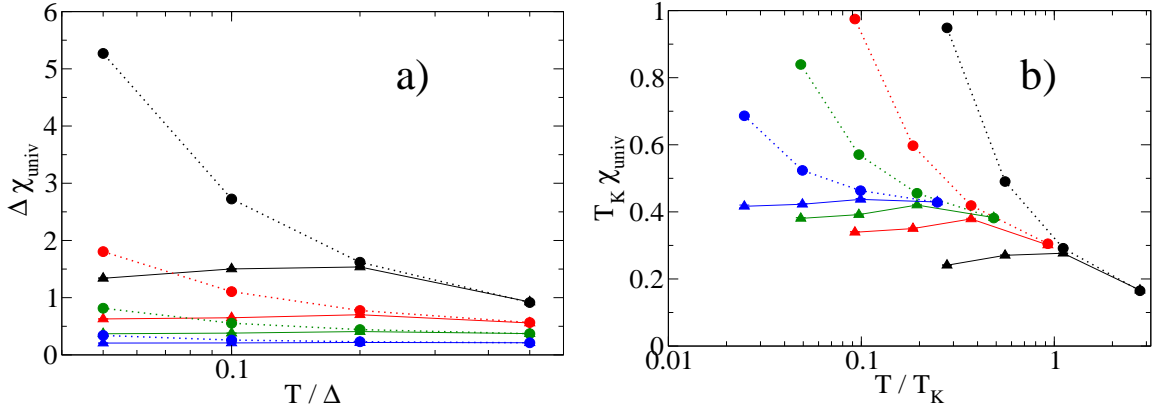}
\caption{(Color online) Rescaled susceptibility $\chi_{\rm univ}$ in the grand canonical ensemble. Same data as in Fig.\,\ref{fig:univsus_gc}, but now plotted for fixed values of $T_{\rm K}/\Delta_\rmR$, i.e. a fixed Hamiltonian. Points connected by solid (dotted) lines have $161$ ($160$) sites in the bath, resulting in an odd (even) value for $\langle N \rangle$. Results are plotted in units of (a) $\Delta_\rmR$ and (b) $T_{\rm K}$. Magnitude of coupling: from top to bottom in (a) and outside to inside in (b), $\mathcal{J}\!=\!0.19,\,0.23,\,0.26$, and $0.30$; these correspond to $T_{\rm K}/\Delta_\rmR\!=\!0.18,\,0.54,\,1.03$ and $2.02$ ($T_{\rm K}$ used here determined as in Fig.\,\ref{fig:univsus_gc}).}
\label{fig:fixedH1} 
\end{figure*}

Fig.\,\ref{fig:EST} shows the results of our calculations, plotted as a function of $T_{\rm K}/\Delta_\rmR$. We emphasize three features: (i) The inset shows that the fit of our QMC data to the simple two-level singlet-triplet form is indeed very good. (ii) The limiting behaviors of Eq.\,(\ref{eq:FST}) are clearly seen. (iii) Data for a wide variety of bare parameters is shown; the excellent collapse onto a single curve in the main figure is a clear demonstration of the expected universality.

We now turn to the local susceptibility and explore the expected scaling ansatz
\begin{equation}
\label{eq:univ_ansatz}
z_{\chi_{\rm loc}}^{-1} T_{\rm K}\chi_{\rm loc}= 
f\left(\frac{{T}}{\Delta_\rmR},\frac{{T}}{T_{\rm K}}\right) \;;
\end{equation}
for variety, we use the grand canonical ensemble. By fixing the chemical potential in the middle of the spectrum of the reservoir, particle-hole symmetry ensures that even in the presence of the Kondo coupling the mean number of particles $\langle N\rangle$ has a fixed parity: if $\mu$ is aligned with a level, $\langle N\rangle$ is odd, while if $\mu$ falls exactly between two levels, $\langle N\rangle$ is even.

Figs.\,\ref{fig:univsus_gc} and \ref{fig:fixedH1} show our QMC results. In Fig.\,\ref{fig:univsus_gc}, we demonstrate the expected scaling by showing the susceptibility as a function of $T/T_{\rm K}$ for fixed $T/\Delta_\rmR$ [ie. ``slices'' of the function $f$ in (\ref{eq:univ_ansatz}) are shown]. A wide variety of bare parameters are used in order to map out the full cross over, and a good data collapse is found. Note that in the low temperature limit, the universal function for $\langle N\rangle$ even is substantially different from that for $\langle N\rangle$ odd.

One technical point concerning the determination of $T_{\rm K}$:
We expect that $T_{\rm K}$ is not affected by finite size effects (in the absence of mesoscopic fluctuations) -- it is determined by the mean density $\rho$, the bandwidth $D$, and the coupling $J$ in the same way as the ``bulk problem''. Indeed, this is explicitly verified in Refs.\,\onlinecite{KaulEPL05} and\,\onlinecite{Yoo05}. However, for practical numerical purposes the 2-loop perturbative formula $T_{\rm K}^{\rm pert}\!=\!D\sqrt{J\rho}\,e^{-1/J\rho}$ works only approximately for $J\rho \!\agt\! 0.20$, and so we need to fit $T_{\rm K}$ numerically in this regime. Since the overall scale of the Kondo temperature is arbitrary, we set $T^{\rm fit}_{\rm K}\!=\!T^{\rm pert}_{\rm K}$ for $J\rho\!=\!0.19$. Having fixed this value, we can then obtain $T^{\rm fit}_{\rm K}$ for all other couplings by choosing it to get the best data collapse in Fig.\,\ref{fig:univsus_gc}. If we were truly in the perturbative regime, this extracted value would always agree well with the perturbative formula. The comparison between $T^{\rm fit}_{\rm K}$ and $T_{\rm K}^{\rm pert}$ is made in the inset of Fig.\,\ref{fig:univsus_gc}. As expected, for stronger couplings there are deviations from the perturbative value. After this fit, there are no further free parameters used to analyze the data.

While the way the data is presented in Fig.\,\ref{fig:univsus_gc} is convenient for demonstrating the universal scaling Eq.\,(\ref{eq:univ_ansatz}), it is more natural to present the data for fixed values of $T_{\rm K}/\Delta_\rmR$ since this corresponds to a fixed geometry or Hamiltonian. This is done in Fig.\,\ref{fig:fixedH1}. The curves for odd and even $\langle N\rangle$ agree for $T/\Delta_\rmR \!\ge\! 0.2$. For lower temperature, the susceptibility saturates for $\langle N\rangle$ odd, while it shows the expected Curie law for $\langle N\rangle$ even. When scaled by $T_{\rm K}$ [panel (b)], the curves for both parities follow the bulk universal curve as temperature decreases until they separate from each other when $T/\Delta_\rmR \!\approx\! 0.2$.

\begin{figure*}
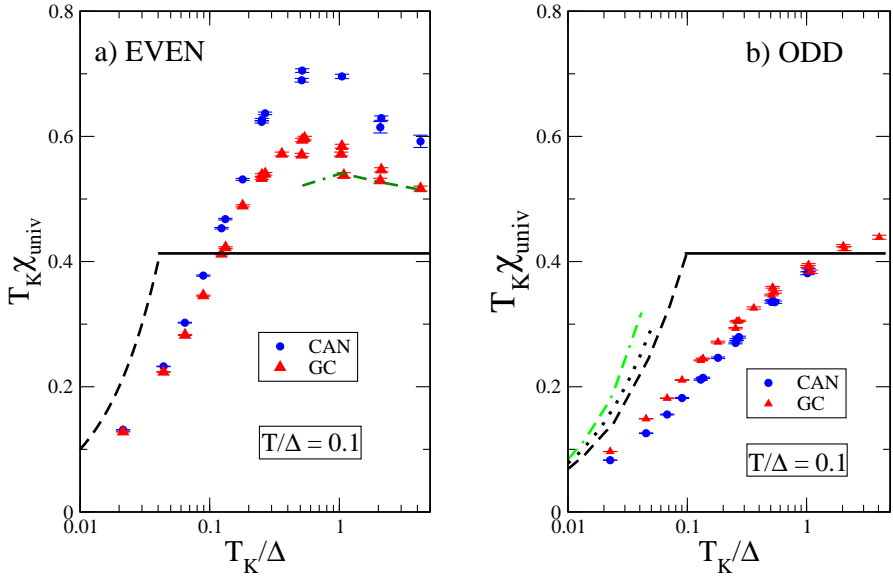

\includegraphics[height=3.in]{can_even_2009-05.eps} \hspace{0.6cm}
\includegraphics[height=3.in]{can_odd_2009-05.eps}
\caption{(Color online) Comparison of canonical and grand-canonical ensemble results for the rescaled susceptibility 
$\chi_{\rm univ} \!=\! z^{-1}_{\rm loc} \chi_{\rm loc}$ [in units of
$(g \mu_B)^2/4$] at the fixed temperature $T/\Delta_\rmR\!=\!0.1$ for an
(a) even and (b) odd number of particles. Symbols: Quantum Monte Carlo
results for the canonical (circles) and grand-canonical (triangles)
ensembles. Solid line: Bulk zero temperature limit. Dashed lines: Weak
coupling expression Eq.\,(\protect{\ref{eq:chi_loc_low}}); for odd
case, uses QMC values of $\delta E_{\rm ST}$. Doted line: Weak
coupling expression Eq.\,(\protect{\ref{eq:chi_loc_low}}) using the
asymptotic expression (\ref{eq:FST}) for $\delta E_{\rm
  ST}$. Dash-dotted lines: Average of the even and odd canonical
results to compare with the grand-canonical result [using the
numerical quantum Monte Carlo data in (a) but the asymptotic
expressions Eq.\,(\protect{\ref{eq:chi_loc_low}}) in (b)]. 
}
\label{fig:can_eo} 
\end{figure*}

\subsection{Limiting regimes and interpolations for $T \!\ll\! \Delta_\rmR$}

To close this section, we come back to the limiting behaviors of the susceptibility discussed earlier. We show in detail how the interpolation between them takes place, as well as the relation between the results for the canonical and grand-canonical ensembles. Fig.\,\ref{fig:can_eo} shows our QMC results yet another way: for fixed $T/\Delta_\rmR \!=\!0.1$ but over the whole transition regime of $T_{\rm K}/\Delta_\rmR$ and for both ensembles. As expected, the results from the canonical ensemble also satisfy the scaling ansatz. We see moreover that the weak and strong coupling limits of the canonical QMC data are reasonable: it is in good agreement with the expressions Eq.\,(\ref{eq:chi_loc_low}) for the weak coupling regime, and tends to the bulk zero temperature limit $f_\chi(0)$ for large $T_{\rm K}/\Delta_\rmR$. (Note, however, that the numerics were not pushed as deep into the limiting regimes as for $\delta E_{\rm ST}$ in Fig.\,\ref{fig:EST}).

The strong dependence of the cross-over behavior on the parity is unexpected. The odd case has a featureless transition and reaches the large $T_{\rm K}/\Delta_\rmR$ limit from below. In contrast, the susceptibility in the even case overshoots the large $T_{\rm K}/\Delta_\rmR$ value and then approaches $f_\chi(0)$ from above. As the even and odd canonical results are so different, this provides an excellent test of the connection between the canonical and grand-canonical results discussed in Section \ref{sec:grand-canonical}: the grand-canonical result in the even case agrees with the average of the two canonical results (green line). As a consequence, the grand-canonical susceptibility for even $\langle N \rangle$ differs considerably from the even $N$ canonical result even for fairly large $T_{\rm K}/\Delta_\rmR$. 

For weak coupling, although the two expressions in Eq.\,(\ref{eq:chi_loc_low}) appear quite different, they lead to similar numerical values for a very large range of $T/\Delta_\rmR$, and in particular for the value $0.1$ used in Fig.\,\ref{fig:can_eo}. As a consequence, the canonical and grand-canonical data in the odd case do not differ very much in the weak coupling regime. Note however that this is somewhat emphasized here by the fact that $T_{\rm K} \chi_{\rm loc}$ is plotted rather than simply the susceptibility itself.

\section{Conclusions}
\label{sec:conclusions}

Our focus in this paper has been on the many-body spectrum of a finite size Kondo system. We have in mind a ``magnetic impurity'' -- either a real one or an effective one formed by a small quantum dot -- coupled to a finite fermionic reservoir (Fig.\,\ref{fig:setup}). Such a system can certainly be made with current technology; indeed, using quantum dots, a \textit{tunable} connection between the small dot (S) and reservoir (R) could be made, allowing a direct investigation of the parametric evolution of properties as function of the coupling. The emphasis throughout the paper is on experimentally observable consequences.

We start with a theorem for the ground state spin of the combined R-S system. Because a crossing of the ground state is forbidden, this can be obtained from simple perturbation theory -- the result for different cases is given in Table I. The theorem is a straightforward extension of the classic theorems of Mattis \cite{Mattis67} and Marshall \cite{Marshall55}. 

A schematic picture of the spectrum of low-lying states as a function of the coupling $J$ can be constructed using perturbation theory and plausibility arguments. Figs.\,\ref{fig:illust} and \ref{fig:illust_us} show results for the screened and under-screened cases, respectively. In this respect we are greatly aided by having a perturbation theory available not only at weak coupling but also at strong coupling  (Nozi\`eres' Fermi liquid theory).

The first observable property that we focus on is the nonlinear $I$-$V$ curve of such an R-S system. Using a rate equation approach, we find the differential conductance as a function of the bias voltage for a number of cases (Figs.\,\ref{fig:ass1} and \ref{fig:ass2}), with certain simplifying assumptions so that the identification of the different transitions is clear.
The key result is that the splitting with magnetic field combined with the magnitude of $dI/dV$ can be used to deduce the spin of the low-lying excited states. Thus an experiment could obtain the information needed to compare with the theoretical schematic picture.

The second observable that we treat is the magnetic susceptibility; in the low temperature limit, this is simply related to the low-lying states of the system. We study the general behavior of both the impurity and local susceptibility, finding that they are markedly different. An extensive example illustrates the general features: A quantum Monte Carlo calculation yields results for the singlet-triplet energy gap (impurity susceptibility) and the local susceptibility in the ``clean Kondo box'' model in which the levels are equally spaced and all levels couple to the impurity with the same amplitude. Results in Figs.\,\ref{fig:EST}-\ref{fig:can_eo} show the expected universality in this model, the strong even-odd effects, and the difference between using the canonical and grand-canonical ensemble.

Clearly the finite size spectrum of engineered many-body systems is a rich area for future experiments. We hope that our schematic arguments plus results for two observable properties will persuade researchers to undertake them. 


\acknowledgments

We thank J. Yoo and P. Simon for useful discussions.
This work was supported in part by the U.S.\ NSF Grant No.\ DMR-0506953 and Hungarian Grants OTKA Nos.\ T046267, T046303, and NF061726. D.U.\ and H.U.B.\ thank the Aspen Center for Physics for its hospitality.

\appendix

\section{Marshall's Theorem}
\label{appendix:marthm}

\textit{Theorem:} If a Hermitian matrix $H$ and a basis set 
$|\phi_\alpha \rangle$ satisfy two conditions,\\
(i) $\langle \phi_\alpha|H|\phi_\beta\rangle < 0$ for
$\alpha \neq \beta$, and\\
(ii) for every pair $\alpha$ and $\beta$, 
$\langle \phi_\alpha|H^n|\phi_\beta\rangle \neq 0$ for at least one integer $n$, \\ 
then the ground state of $H$ is unique (i.e. non-degenerate).

We prove the theorem in a few steps:

\textit{Step I:} First, we prove that if the conditions above are
satisfied, then the following simple result is true: if $\sum_{\alpha}
f_\alpha | \phi_\alpha \rangle$ is a ground state, then so is $\sum_{\alpha} |f_\alpha| | \phi_\alpha \rangle$. 
To show this, consider the following ``Perron-Frobenius'' inequality:
\begin{equation}
\sum_{\alpha, \beta} |f_\alpha|\, |f_\beta| \langle \phi_\alpha | H |
\phi_\beta \rangle \leq \sum_{\alpha, \beta} f_\alpha f_\beta \langle
\phi_\alpha | H | \phi_\beta \rangle  \;.
\end{equation}
Note that the two sides of the inequality differ only for terms with $\alpha \!\neq\! \beta$. All such terms on the left-hand side are negative because of condition (i) above, while on the right-hand side the sign may be positive, depending on the sign of $f_\alpha$; hence, the inequality holds. Physically, the right and left side of the inequality are, in fact, the expectation values of the energy in the ground state $\sum_\alpha f_\alpha |\phi_\alpha \rangle$ and in $\sum_\alpha |f_\alpha|\, |\phi_\alpha \rangle$, respectively. Since the expectation value of the Hamiltonian in the ground state is always the lowest, $\sum_\alpha |f_\alpha|\, |\phi_\alpha \rangle$ must also be a ground state.

\textit{Step II:} We now show that any ground state described by
$f_\alpha$ must have $f_\alpha \geq 0$. For convenience we separate
$H=H_{\rm D}+H_{\rm OD}$, where $H_{\rm D}$ and $H_{\rm OD}$ are the
diagonal and off-diagonal parts of $H$ with respect to the basis
$|\phi_\alpha \rangle$. In this language the Schr\"odinger equation expressed for the states $\sum_\alpha f_\alpha |\phi_\alpha \rangle$ and $\sum_\alpha |f_\alpha|\, |\phi_\alpha \rangle$ becomes
\begin{eqnarray}
\label{marshall:eq:sch_eqn}
\langle \phi_\alpha |H_{\rm D}| \phi_\alpha \rangle f_\alpha 
- \sum_{\beta}|\langle \phi_\alpha |H_{\rm OD}| \phi_\beta
\rangle|f_\beta = E_{\rm G} f_\alpha \quad\\ 
\langle \phi_\alpha |H_{\rm D}| \phi_\alpha \rangle |f_\alpha| 
- \sum_{\beta}|\langle \phi_\alpha |H_{\rm OD}| \phi_\beta
\rangle|\, |f_\beta| = E_{\rm G} |f_\alpha| \quad
\end{eqnarray}
where $E_{\rm G}$ is the ground state energy. These two equations combine to give
\begin{equation}
\left|\big(\langle \phi_\alpha |H_{\rm D}| \phi_\alpha \rangle - E_{\rm G}\big) f_\alpha \right| = 
\big|\langle \phi_\alpha |H_{\rm D}| \phi_\alpha \rangle - E_{\rm G}\big|\, | f_\alpha |.
\end{equation}
Note, however, that $\langle \phi_\alpha |H_{\rm D}| \phi_\alpha
\rangle - E_{\rm G}\geq\! 0$, because $|\phi_\alpha \rangle$ cannot have a lower energy expectation value than the ground state. Thus, $f_{\alpha}\geq 0$ follows.

\textit{Step III:} The next step is to use condition (ii) from the theorem to show that the stricter condition $f_\alpha \!>\! 0$ holds. This is most easily seen in Eq.\,(\ref{marshall:eq:sch_eqn}): if $f_{\alpha_1}\!=\!0$ for one $\alpha_1$, then, using condition (ii), all the $f_\alpha$ must be zero. Since this cannot be the case, all $f_\alpha \!\neq\! 0$. Thus, combining with the result of \textit{Step II}, we conclude that $f_{\alpha}\!>\!0$.

\textit{Step IV:} We have thus shown that every ground state of $H$ has a positive definite expansion in the basis set $|\phi_\alpha \rangle$. Since there cannot be two simultaneously orthogonal \textit{and} positive definite vectors, the ground state of $H$ must be non-degenerate. This proves the theorem.

\bibliography{general_ref,kondo,nano,footnotes}

\begin{thebibliography}{63}
\expandafter\ifx\csname natexlab\endcsname\relax\def\natexlab#1{#1}\fi
\expandafter\ifx\csname bibnamefont\endcsname\relax
  \def\bibnamefont#1{#1}\fi
\expandafter\ifx\csname bibfnamefont\endcsname\relax
  \def\bibfnamefont#1{#1}\fi
\expandafter\ifx\csname citenamefont\endcsname\relax
  \def\citenamefont#1{#1}\fi
\expandafter\ifx\csname url\endcsname\relax
  \def\url#1{\texttt{#1}}\fi
\expandafter\ifx\csname urlprefix\endcsname\relax\def\urlprefix{URL }\fi
\providecommand{\bibinfo}[2]{#2}
\providecommand{\eprint}[2][]{\url{#2}}

\bibitem[{\citenamefont{Hewson}(1993)}]{HewsonBook}
\bibinfo{author}{\bibfnamefont{A.}~\bibnamefont{Hewson}},
  \emph{\bibinfo{title}{The Kondo Problem to Heavy Fermions}}
  (\bibinfo{publisher}{Cambridge University Press},
  \bibinfo{address}{Cambridge}, \bibinfo{year}{1993}).

\bibitem[{\citenamefont{Glazman and Raikh}(1988)}]{Glazman88}
\bibinfo{author}{\bibfnamefont{L.}~\bibnamefont{Glazman}} \bibnamefont{and}
  \bibinfo{author}{\bibfnamefont{M.}~\bibnamefont{Raikh}},
  \bibinfo{journal}{Pis'ma Zh. Eksp. Teor. Fiz.} \textbf{\bibinfo{volume}{47}},
  \bibinfo{pages}{378} (\bibinfo{year}{1988}), \bibinfo{note}{[JETP Lett. {\bf
  47} (1988) 452]}.

\bibitem[{\citenamefont{Ng and Lee}(1988)}]{Ng88}
\bibinfo{author}{\bibfnamefont{T.~K.} \bibnamefont{Ng}} \bibnamefont{and}
  \bibinfo{author}{\bibfnamefont{P.~A.} \bibnamefont{Lee}},
  \bibinfo{journal}{Phys. Rev. Lett.} \textbf{\bibinfo{volume}{61}},
  \bibinfo{pages}{1768} (\bibinfo{year}{1988}).

\bibitem[{\citenamefont{Glazman and Pustilnik}(2005)}]{GlazmanHouches05}
\bibinfo{author}{\bibfnamefont{L.}~\bibnamefont{Glazman}} \bibnamefont{and}
  \bibinfo{author}{\bibfnamefont{M.}~\bibnamefont{Pustilnik}}, in
  \emph{\bibinfo{booktitle}{Nanophysics: Coherence and Transport}}, edited by
  \bibinfo{editor}{\bibfnamefont{H.}~\bibnamefont{Bouchiat}},
  \bibinfo{editor}{\bibfnamefont{Y.}~\bibnamefont{Gefen}},
  \bibinfo{editor}{\bibfnamefont{S.}~\bibnamefont{Gueron}},
  \bibinfo{editor}{\bibfnamefont{G.}~\bibnamefont{Montambaux}},
  \bibnamefont{and} \bibinfo{editor}{\bibfnamefont{J.}~\bibnamefont{Dalibard}}
  (\bibinfo{publisher}{Elsevier}, \bibinfo{year}{2005}), pp.
  \bibinfo{pages}{427--478}.

\bibitem[{\citenamefont{Zar{\'a}nd}(2006)}]{ZarandRev06}
\bibinfo{author}{\bibfnamefont{G.}~\bibnamefont{Zar{\'a}nd}},
  \bibinfo{journal}{Phil. Mag.} \textbf{\bibinfo{volume}{86}},
  \bibinfo{pages}{2043} (\bibinfo{year}{2006}).

\bibitem[{\citenamefont{Grobis et~al.}(2007)\citenamefont{Grobis, Rau, Potok,
  and Goldhaber-Gordon}}]{GoldhaberGRev07}
\bibinfo{author}{\bibfnamefont{M.}~\bibnamefont{Grobis}},
  \bibinfo{author}{\bibfnamefont{I.~G.} \bibnamefont{Rau}},
  \bibinfo{author}{\bibfnamefont{R.~M.} \bibnamefont{Potok}}, \bibnamefont{and}
  \bibinfo{author}{\bibfnamefont{D.}~\bibnamefont{Goldhaber-Gordon}}, in
  \emph{\bibinfo{booktitle}{Handbook of Magnetism and Advanced Magnetic
  Materials, Vol. 5}}, edited by
  \bibinfo{editor}{\bibfnamefont{H.}~\bibnamefont{Kronm{\"u}ller}}
  \bibnamefont{and} \bibinfo{editor}{\bibfnamefont{S.}~\bibnamefont{Parkin}}
  (\bibinfo{publisher}{Wiley}, \bibinfo{address}{New York},
  \bibinfo{year}{2007}).

\bibitem[{\citenamefont{Goldhaber-Gordon
  et~al.}(1998)\citenamefont{Goldhaber-Gordon, Shtrikman, Mahalu,
  Abusch-Magder, Meirav, and Kastner}}]{Goldhaber98}
\bibinfo{author}{\bibfnamefont{D.}~\bibnamefont{Goldhaber-Gordon}},
  \bibinfo{author}{\bibfnamefont{H.}~\bibnamefont{Shtrikman}},
  \bibinfo{author}{\bibfnamefont{D.}~\bibnamefont{Mahalu}},
  \bibinfo{author}{\bibfnamefont{D.}~\bibnamefont{Abusch-Magder}},
  \bibinfo{author}{\bibfnamefont{U.}~\bibnamefont{Meirav}}, \bibnamefont{and}
  \bibinfo{author}{\bibfnamefont{M.}~\bibnamefont{Kastner}},
  \bibinfo{journal}{Nature} \textbf{\bibinfo{volume}{391}},
  \bibinfo{pages}{156} (\bibinfo{year}{1998}).

\bibitem[{\citenamefont{Cronenwett et~al.}(1998)\citenamefont{Cronenwett,
  Oosterkamp, and Kouwenhoven}}]{Cronenwett98}
\bibinfo{author}{\bibfnamefont{S.~M.} \bibnamefont{Cronenwett}},
  \bibinfo{author}{\bibfnamefont{T.~H.} \bibnamefont{Oosterkamp}},
  \bibnamefont{and} \bibinfo{author}{\bibfnamefont{L.~P.}
  \bibnamefont{Kouwenhoven}}, \bibinfo{journal}{Science}
  \textbf{\bibinfo{volume}{281}}, \bibinfo{pages}{540} (\bibinfo{year}{1998}).

\bibitem[{\citenamefont{Pustilnik et~al.}(2001)\citenamefont{Pustilnik,
  Glazman, Cobden, and Kouwenhoven}}]{Pustilnik01}
\bibinfo{author}{\bibfnamefont{M.}~\bibnamefont{Pustilnik}},
  \bibinfo{author}{\bibfnamefont{L.~I.} \bibnamefont{Glazman}},
  \bibinfo{author}{\bibfnamefont{D.~H.} \bibnamefont{Cobden}},
  \bibnamefont{and} \bibinfo{author}{\bibfnamefont{L.~P.}
  \bibnamefont{Kouwenhoven}}, \bibinfo{journal}{Lecture notes in Physics}
  \textbf{\bibinfo{volume}{3}}, \bibinfo{pages}{579} (\bibinfo{year}{2001}),
  \bibinfo{note}{cond-mat/0010336}.

\bibitem[{\citenamefont{Oreg and Goldhaber-Gordon}(2003)}]{Oreg03}
\bibinfo{author}{\bibfnamefont{Y.}~\bibnamefont{Oreg}} \bibnamefont{and}
  \bibinfo{author}{\bibfnamefont{D.}~\bibnamefont{Goldhaber-Gordon}},
  \bibinfo{journal}{Phys. Rev. Lett.} \textbf{\bibinfo{volume}{90}},
  \bibinfo{pages}{136602} (\bibinfo{year}{2003}).

\bibitem[{\citenamefont{Potok et~al.}(2007)\citenamefont{Potok, Rau, Shtrikman,
  Oreg, and Goldhaber-Gordon}}]{Potok06}
\bibinfo{author}{\bibfnamefont{R.~M.} \bibnamefont{Potok}},
  \bibinfo{author}{\bibfnamefont{I.~G.} \bibnamefont{Rau}},
  \bibinfo{author}{\bibfnamefont{H.}~\bibnamefont{Shtrikman}},
  \bibinfo{author}{\bibfnamefont{Y.}~\bibnamefont{Oreg}}, \bibnamefont{and}
  \bibinfo{author}{\bibfnamefont{D.}~\bibnamefont{Goldhaber-Gordon}},
  \bibinfo{journal}{Nature} \textbf{\bibinfo{volume}{446}},
  \bibinfo{pages}{167} (\bibinfo{year}{2007}).

\bibitem[{\citenamefont{Borda et~al.}(2003)\citenamefont{Borda, Zar{\'a}nd,
  Hofstetter, Halperin, and von Delft}}]{Borda03}
\bibinfo{author}{\bibfnamefont{L.}~\bibnamefont{Borda}},
  \bibinfo{author}{\bibfnamefont{G.}~\bibnamefont{Zar{\'a}nd}},
  \bibinfo{author}{\bibfnamefont{W.}~\bibnamefont{Hofstetter}},
  \bibinfo{author}{\bibfnamefont{B.~I.} \bibnamefont{Halperin}},
  \bibnamefont{and} \bibinfo{author}{\bibfnamefont{J.}~\bibnamefont{von
  Delft}}, \bibinfo{journal}{Phys. Rev. Lett.} \textbf{\bibinfo{volume}{90}},
  \bibinfo{pages}{026602} (\bibinfo{year}{2003}).

\bibitem[{\citenamefont{Le~Hur and Simon}(2003)}]{LeHur03}
\bibinfo{author}{\bibfnamefont{K.}~\bibnamefont{Le~Hur}} \bibnamefont{and}
  \bibinfo{author}{\bibfnamefont{P.}~\bibnamefont{Simon}},
  \bibinfo{journal}{Phys. Rev. B} \textbf{\bibinfo{volume}{67}},
  \bibinfo{pages}{201308} (\bibinfo{year}{2003}).

\bibitem[{\citenamefont{Le~Hur et~al.}(2004)\citenamefont{Le~Hur, Simon, and
  Borda}}]{LeHur04}
\bibinfo{author}{\bibfnamefont{K.}~\bibnamefont{Le~Hur}},
  \bibinfo{author}{\bibfnamefont{P.}~\bibnamefont{Simon}}, \bibnamefont{and}
  \bibinfo{author}{\bibfnamefont{L.}~\bibnamefont{Borda}},
  \bibinfo{journal}{Phys. Rev. B} \textbf{\bibinfo{volume}{69}},
  \bibinfo{pages}{045326} (\bibinfo{year}{2004}).

\bibitem[{\citenamefont{Galpin et~al.}(2005)\citenamefont{Galpin, Logan, and
  Krishnamurthy}}]{Galpin05}
\bibinfo{author}{\bibfnamefont{M.~R.} \bibnamefont{Galpin}},
  \bibinfo{author}{\bibfnamefont{D.~E.} \bibnamefont{Logan}}, \bibnamefont{and}
  \bibinfo{author}{\bibfnamefont{H.~R.} \bibnamefont{Krishnamurthy}},
  \bibinfo{journal}{Phy. Rev. Lett.} \textbf{\bibinfo{volume}{94}},
  \bibinfo{pages}{186406} (\bibinfo{year}{2005}).

\bibitem[{\citenamefont{Le~Hur et~al.}(2007)\citenamefont{Le~Hur, Simon, and
  Loss}}]{LeHur07}
\bibinfo{author}{\bibfnamefont{K.}~\bibnamefont{Le~Hur}},
  \bibinfo{author}{\bibfnamefont{P.}~\bibnamefont{Simon}}, \bibnamefont{and}
  \bibinfo{author}{\bibfnamefont{D.}~\bibnamefont{Loss}},
  \bibinfo{journal}{Phy. Rev. B} \textbf{\bibinfo{volume}{75}},
  \bibinfo{pages}{035332} (\bibinfo{year}{2007}).

\bibitem[{\citenamefont{Choi et~al.}(2005)\citenamefont{Choi, Lopez, and
  Aguado}}]{Choi05}
\bibinfo{author}{\bibfnamefont{M.-S.} \bibnamefont{Choi}},
  \bibinfo{author}{\bibfnamefont{R.}~\bibnamefont{Lopez}}, \bibnamefont{and}
  \bibinfo{author}{\bibfnamefont{R.}~\bibnamefont{Aguado}},
  \bibinfo{journal}{Phys. Rev. Lett.} \textbf{\bibinfo{volume}{95}},
  \bibinfo{pages}{067204} (\bibinfo{year}{2005}).

\bibitem[{\citenamefont{Makarovski
  et~al.}(2007{\natexlab{a}})\citenamefont{Makarovski, Zhukov, Liu, and
  Finkelstein}}]{Makarovski07a}
\bibinfo{author}{\bibfnamefont{A.}~\bibnamefont{Makarovski}},
  \bibinfo{author}{\bibfnamefont{A.}~\bibnamefont{Zhukov}},
  \bibinfo{author}{\bibfnamefont{J.}~\bibnamefont{Liu}}, \bibnamefont{and}
  \bibinfo{author}{\bibfnamefont{G.}~\bibnamefont{Finkelstein}},
  \bibinfo{journal}{Phys. Rev. B} \textbf{\bibinfo{volume}{75}},
  \bibinfo{pages}{241407} (\bibinfo{year}{2007}{\natexlab{a}}).

\bibitem[{\citenamefont{Makarovski
  et~al.}(2007{\natexlab{b}})\citenamefont{Makarovski, Liu, and
  Finkelstein}}]{Makarovski07b}
\bibinfo{author}{\bibfnamefont{A.}~\bibnamefont{Makarovski}},
  \bibinfo{author}{\bibfnamefont{J.}~\bibnamefont{Liu}}, \bibnamefont{and}
  \bibinfo{author}{\bibfnamefont{G.}~\bibnamefont{Finkelstein}},
  \bibinfo{journal}{Phys. Rev. Lett.} \textbf{\bibinfo{volume}{99}},
  \bibinfo{pages}{066801} (\bibinfo{year}{2007}{\natexlab{b}}).

\bibitem[{\citenamefont{Kouwenhoven et~al.}(1997)\citenamefont{Kouwenhoven,
  Marcus, McEuen, Tarucha, Wetervelt, and Wingreen}}]{Kouwenhoven97}
\bibinfo{author}{\bibfnamefont{L.~P.} \bibnamefont{Kouwenhoven}},
  \bibinfo{author}{\bibfnamefont{C.~M.} \bibnamefont{Marcus}},
  \bibinfo{author}{\bibfnamefont{P.~L.} \bibnamefont{McEuen}},
  \bibinfo{author}{\bibfnamefont{S.}~\bibnamefont{Tarucha}},
  \bibinfo{author}{\bibfnamefont{R.~M.} \bibnamefont{Wetervelt}},
  \bibnamefont{and} \bibinfo{author}{\bibfnamefont{N.~S.}
  \bibnamefont{Wingreen}}, in \emph{\bibinfo{booktitle}{Mesoscopic Electron
  Transport}}, edited by \bibinfo{editor}{\bibfnamefont{L.~L.}
  \bibnamefont{Sohn}},
  \bibinfo{editor}{\bibfnamefont{G.}~\bibnamefont{Sch{\"o}n}},
  \bibnamefont{and} \bibinfo{editor}{\bibfnamefont{L.~P.}
  \bibnamefont{Kouwenhoven}} (\bibinfo{publisher}{Kluwer},
  \bibinfo{address}{Dordrecht}, \bibinfo{year}{1997}), pp.
  \bibinfo{pages}{105--214}.

\bibitem[{\citenamefont{Akkermans and Montambaux}(2007)}]{AkkerMontamBook}
\bibinfo{author}{\bibfnamefont{E.}~\bibnamefont{Akkermans}} \bibnamefont{and}
  \bibinfo{author}{\bibfnamefont{G.}~\bibnamefont{Montambaux}},
  \emph{\bibinfo{title}{Mesoscopic Physics of Electrons and Phonons}}
  (\bibinfo{publisher}{Cambridge University Press},
  \bibinfo{address}{Cambridge, UK}, \bibinfo{year}{2007}).

\bibitem[{\citenamefont{Argaman et~al.}(1993)\citenamefont{Argaman, Imry, and
  Smilansky}}]{Argaman93}
\bibinfo{author}{\bibfnamefont{N.}~\bibnamefont{Argaman}},
  \bibinfo{author}{\bibfnamefont{Y.}~\bibnamefont{Imry}}, \bibnamefont{and}
  \bibinfo{author}{\bibfnamefont{U.}~\bibnamefont{Smilansky}},
  \bibinfo{journal}{Phys. Rev. B} \textbf{\bibinfo{volume}{47}},
  \bibinfo{pages}{4440} (\bibinfo{year}{1993}).

\bibitem[{\citenamefont{Zar{\'a}nd and Udvardi}(1996)}]{Zarand96}
\bibinfo{author}{\bibfnamefont{G.}~\bibnamefont{Zar{\'a}nd}} \bibnamefont{and}
  \bibinfo{author}{\bibfnamefont{L.}~\bibnamefont{Udvardi}},
  \bibinfo{journal}{Phys. Rev. B} \textbf{\bibinfo{volume}{54}},
  \bibinfo{pages}{7606} (\bibinfo{year}{1996}).

\bibitem[{\citenamefont{Kettemann}(2004)}]{Kettemann04}
\bibinfo{author}{\bibfnamefont{S.}~\bibnamefont{Kettemann}}, in
  \emph{\bibinfo{booktitle}{Quantum Information and Decoherence in
  Nanosystems}}, edited by \bibinfo{editor}{\bibfnamefont{D.~C.}
  \bibnamefont{Glattli}},
  \bibinfo{editor}{\bibfnamefont{M.}~\bibnamefont{Sanquer}}, \bibnamefont{and}
  \bibinfo{editor}{\bibfnamefont{J.~T.~T.} \bibnamefont{Van}}
  (\bibinfo{publisher}{The Gioi Publishers}, \bibinfo{year}{2004}), p.
  \bibinfo{pages}{259}, \bibinfo{note}{(cond-mat/0409317)}.

\bibitem[{\citenamefont{Kaul et~al.}(2005)\citenamefont{Kaul, Ullmo,
  Chandrasekharan, and Baranger}}]{KaulEPL05}
\bibinfo{author}{\bibfnamefont{R.~K.} \bibnamefont{Kaul}},
  \bibinfo{author}{\bibfnamefont{D.}~\bibnamefont{Ullmo}},
  \bibinfo{author}{\bibfnamefont{S.}~\bibnamefont{Chandrasekharan}},
  \bibnamefont{and} \bibinfo{author}{\bibfnamefont{H.~U.}
  \bibnamefont{Baranger}}, \bibinfo{journal}{Europhys. Lett.}
  \textbf{\bibinfo{volume}{71}}, \bibinfo{pages}{973} (\bibinfo{year}{2005}).

\bibitem[{\citenamefont{Yoo et~al.}(2005)\citenamefont{Yoo, Chandrasekharan,
  Kaul, Ullmo, and Baranger}}]{Yoo05}
\bibinfo{author}{\bibfnamefont{J.}~\bibnamefont{Yoo}},
  \bibinfo{author}{\bibfnamefont{S.}~\bibnamefont{Chandrasekharan}},
  \bibinfo{author}{\bibfnamefont{R.~K.} \bibnamefont{Kaul}},
  \bibinfo{author}{\bibfnamefont{D.}~\bibnamefont{Ullmo}}, \bibnamefont{and}
  \bibinfo{author}{\bibfnamefont{H.~U.} \bibnamefont{Baranger}},
  \bibinfo{journal}{Phys. Rev. B} \textbf{\bibinfo{volume}{71}},
  \bibinfo{pages}{201309(R)} (\bibinfo{year}{2005}).

\bibitem[{\citenamefont{Ullmo}(2008)}]{UllmoRPP08}
\bibinfo{author}{\bibfnamefont{D.}~\bibnamefont{Ullmo}}, \bibinfo{journal}{Rep.
  Prog. Phys.} \textbf{\bibinfo{volume}{71}}, \bibinfo{pages}{026001}
  (\bibinfo{year}{2008}).

\bibitem[{\citenamefont{Kettemann and Mucciolo}(2006)}]{Kettemann06}
\bibinfo{author}{\bibfnamefont{S.}~\bibnamefont{Kettemann}} \bibnamefont{and}
  \bibinfo{author}{\bibfnamefont{E.~R.} \bibnamefont{Mucciolo}},
  \bibinfo{journal}{Pis'ma v ZhETF} \textbf{\bibinfo{volume}{83}},
  \bibinfo{pages}{284} (\bibinfo{year}{2006}), \bibinfo{note}{[JETP Letters
  {\bf 83}, 240 (2006)]}.

\bibitem[{\citenamefont{Kettemann and Mucciolo}(2007)}]{Kettemann07}
\bibinfo{author}{\bibfnamefont{S.}~\bibnamefont{Kettemann}} \bibnamefont{and}
  \bibinfo{author}{\bibfnamefont{E.~R.} \bibnamefont{Mucciolo}},
  \bibinfo{journal}{Phys. Rev. B} \textbf{\bibinfo{volume}{75}},
  \bibinfo{pages}{184407} (\bibinfo{year}{2007}).

\bibitem[{\citenamefont{Zhuravlev et~al.}(2007)\citenamefont{Zhuravlev,
  Zharekeshev, Gorelov, Lichtenstein, Mucciolo, and Kettemann}}]{Zhuravlev08}
\bibinfo{author}{\bibfnamefont{A.}~\bibnamefont{Zhuravlev}},
  \bibinfo{author}{\bibfnamefont{I.}~\bibnamefont{Zharekeshev}},
  \bibinfo{author}{\bibfnamefont{E.}~\bibnamefont{Gorelov}},
  \bibinfo{author}{\bibfnamefont{A.~I.} \bibnamefont{Lichtenstein}},
  \bibinfo{author}{\bibfnamefont{E.~R.} \bibnamefont{Mucciolo}},
  \bibnamefont{and}
  \bibinfo{author}{\bibfnamefont{S.}~\bibnamefont{Kettemann}},
  \bibinfo{journal}{arXiv:0706.3456v1}  (\bibinfo{year}{2007}).

\bibitem[{\citenamefont{Thimm et~al.}(1999)\citenamefont{Thimm, Kroha, and von
  Delft}}]{Thimm99}
\bibinfo{author}{\bibfnamefont{W.~B.} \bibnamefont{Thimm}},
  \bibinfo{author}{\bibfnamefont{J.}~\bibnamefont{Kroha}}, \bibnamefont{and}
  \bibinfo{author}{\bibfnamefont{J.}~\bibnamefont{von Delft}},
  \bibinfo{journal}{Phys. Rev. Lett.} \textbf{\bibinfo{volume}{82}},
  \bibinfo{pages}{2143} (\bibinfo{year}{1999}).

\bibitem[{\citenamefont{Simon and Affleck}(2002)}]{Simon02}
\bibinfo{author}{\bibfnamefont{P.}~\bibnamefont{Simon}} \bibnamefont{and}
  \bibinfo{author}{\bibfnamefont{I.}~\bibnamefont{Affleck}},
  \bibinfo{journal}{Phys. Rev. Lett.} \textbf{\bibinfo{volume}{89}},
  \bibinfo{pages}{206602} (\bibinfo{year}{2002}).

\bibitem[{\citenamefont{Cornaglia and Balseiro}(2003)}]{Cornaglia03}
\bibinfo{author}{\bibfnamefont{P.~S.} \bibnamefont{Cornaglia}}
  \bibnamefont{and} \bibinfo{author}{\bibfnamefont{C.~A.}
  \bibnamefont{Balseiro}}, \bibinfo{journal}{Phys. Rev. Lett.}
  \textbf{\bibinfo{volume}{90}}, \bibinfo{pages}{216801}
  (\bibinfo{year}{2003}).

\bibitem[{\citenamefont{Affleck and Simon}(2001)}]{Affleck01}
\bibinfo{author}{\bibfnamefont{I.}~\bibnamefont{Affleck}} \bibnamefont{and}
  \bibinfo{author}{\bibfnamefont{P.}~\bibnamefont{Simon}},
  \bibinfo{journal}{Phys. Rev. Lett.} \textbf{\bibinfo{volume}{86}},
  \bibinfo{pages}{2854} (\bibinfo{year}{2001}).

\bibitem[{\citenamefont{Simon and Affleck}(2001)}]{Simon01}
\bibinfo{author}{\bibfnamefont{P.}~\bibnamefont{Simon}} \bibnamefont{and}
  \bibinfo{author}{\bibfnamefont{I.}~\bibnamefont{Affleck}},
  \bibinfo{journal}{Phys. Rev. B} \textbf{\bibinfo{volume}{64}},
  \bibinfo{pages}{085308} (\bibinfo{year}{2001}).

\bibitem[{\citenamefont{Lewenkopf and Weidenmuller}(2005)}]{Lewenkopf05}
\bibinfo{author}{\bibfnamefont{C.~H.} \bibnamefont{Lewenkopf}}
  \bibnamefont{and} \bibinfo{author}{\bibfnamefont{H.~A.}
  \bibnamefont{Weidenmuller}}, \bibinfo{journal}{Phys. Rev. B}
  \textbf{\bibinfo{volume}{71}}, \bibinfo{pages}{121309}
  (\bibinfo{year}{2005}).

\bibitem[{\citenamefont{Simon et~al.}(2006)\citenamefont{Simon, Salomez, and
  Feinberg}}]{Simon06}
\bibinfo{author}{\bibfnamefont{P.}~\bibnamefont{Simon}},
  \bibinfo{author}{\bibfnamefont{J.}~\bibnamefont{Salomez}}, \bibnamefont{and}
  \bibinfo{author}{\bibfnamefont{D.}~\bibnamefont{Feinberg}},
  \bibinfo{journal}{Phys. Rev. B} \textbf{\bibinfo{volume}{73}},
  \bibinfo{pages}{205325} (\bibinfo{year}{2006}).

\bibitem[{\citenamefont{Cornaglia and Balseiro}(2002)}]{Cornaglia02a}
\bibinfo{author}{\bibfnamefont{P.~S.} \bibnamefont{Cornaglia}}
  \bibnamefont{and} \bibinfo{author}{\bibfnamefont{C.~A.}
  \bibnamefont{Balseiro}}, \bibinfo{journal}{Phys. Rev. B}
  \textbf{\bibinfo{volume}{66}}, \bibinfo{pages}{115303}
  (\bibinfo{year}{2002}).

\bibitem[{\citenamefont{Kaul et~al.}(2006)\citenamefont{Kaul, Zar{\'a}nd,
  Chandrasekharan, Ullmo, and Baranger}}]{KaulPRL06}
\bibinfo{author}{\bibfnamefont{R.~K.} \bibnamefont{Kaul}},
  \bibinfo{author}{\bibfnamefont{G.}~\bibnamefont{Zar{\'a}nd}},
  \bibinfo{author}{\bibfnamefont{S.}~\bibnamefont{Chandrasekharan}},
  \bibinfo{author}{\bibfnamefont{D.}~\bibnamefont{Ullmo}}, \bibnamefont{and}
  \bibinfo{author}{\bibfnamefont{H.~U.} \bibnamefont{Baranger}},
  \bibinfo{journal}{Phys. Rev. Lett.} \textbf{\bibinfo{volume}{96}},
  \bibinfo{pages}{176802} (\bibinfo{year}{2006}).

\bibitem[{\citenamefont{Pereira et~al.}(2008)\citenamefont{Pereira,
  Laflorencie, Affleck, and Halperin}}]{Pereira08}
\bibinfo{author}{\bibfnamefont{R.~G.} \bibnamefont{Pereira}},
  \bibinfo{author}{\bibfnamefont{N.}~\bibnamefont{Laflorencie}},
  \bibinfo{author}{\bibfnamefont{I.}~\bibnamefont{Affleck}}, \bibnamefont{and}
  \bibinfo{author}{\bibfnamefont{B.~I.} \bibnamefont{Halperin}},
  \bibinfo{journal}{Phys. Rev. B} \textbf{\bibinfo{volume}{77}},
  \bibinfo{pages}{125327} (\bibinfo{year}{2008}).

\bibitem[{\citenamefont{Murthy}(2005)}]{Murthy05}
\bibinfo{author}{\bibfnamefont{G.}~\bibnamefont{Murthy}},
  \bibinfo{journal}{Phys. Rev. Lett.} \textbf{\bibinfo{volume}{94}},
  \bibinfo{pages}{126803} (\bibinfo{year}{2005}).

\bibitem[{\citenamefont{Rotter et~al.}(2008)\citenamefont{Rotter, T{\"{u}}reci,
  Alhassid, and Stone}}]{RotterPRL08}
\bibinfo{author}{\bibfnamefont{S.}~\bibnamefont{Rotter}},
  \bibinfo{author}{\bibfnamefont{H.~E.} \bibnamefont{T{\"{u}}reci}},
  \bibinfo{author}{\bibfnamefont{Y.}~\bibnamefont{Alhassid}}, \bibnamefont{and}
  \bibinfo{author}{\bibfnamefont{A.~D.} \bibnamefont{Stone}},
  \bibinfo{journal}{Phys. Rev. Lett.} \textbf{\bibinfo{volume}{100}},
  \bibinfo{pages}{166601} (\bibinfo{year}{2008}).

\bibitem[{\citenamefont{Mattis}(1967)}]{Mattis67}
\bibinfo{author}{\bibfnamefont{D.~C.} \bibnamefont{Mattis}},
  \bibinfo{journal}{Phys. Rev. Lett.} \textbf{\bibinfo{volume}{19}},
  \bibinfo{pages}{1478} (\bibinfo{year}{1967}).

\bibitem[{\citenamefont{Marshall}(1955)}]{Marshall55}
\bibinfo{author}{\bibfnamefont{W.}~\bibnamefont{Marshall}},
  \bibinfo{journal}{Proc. R. Soc. London Ser. {\bf A}}
  \textbf{\bibinfo{volume}{232}}, \bibinfo{pages}{48} (\bibinfo{year}{1955}).

\bibitem[{\citenamefont{Auerbach}(1994)}]{AuerbachBook}
\bibinfo{author}{\bibfnamefont{A.}~\bibnamefont{Auerbach}},
  \emph{\bibinfo{title}{Interacting Electrons and Quantum Magentism}}
  (\bibinfo{publisher}{Springer-Verlag}, \bibinfo{address}{New York},
  \bibinfo{year}{1994}).

\bibitem[{\citenamefont{Wilson}(1975)}]{WilsonRMP75}
\bibinfo{author}{\bibfnamefont{K.~G.} \bibnamefont{Wilson}},
  \bibinfo{journal}{Rev. Mod. Phys.} \textbf{\bibinfo{volume}{47}},
  \bibinfo{pages}{773} (\bibinfo{year}{1975}).

\bibitem[{\citenamefont{Nozi{\`e}res}(1974)}]{Nozieres74}
\bibinfo{author}{\bibfnamefont{P.}~\bibnamefont{Nozi{\`e}res}},
  \bibinfo{journal}{J. Low Temp. Phys.} \textbf{\bibinfo{volume}{17}},
  \bibinfo{pages}{31} (\bibinfo{year}{1974}).

\bibitem[{\citenamefont{Nozi{\`e}res}(1978)}]{Nozieres78}
\bibinfo{author}{\bibfnamefont{P.}~\bibnamefont{Nozi{\`e}res}},
  \bibinfo{journal}{J. Phys. (Paris)} \textbf{\bibinfo{volume}{39}},
  \bibinfo{pages}{1117} (\bibinfo{year}{1978}).

\bibitem[{\citenamefont{van~der Wiel et~al.}(2002)\citenamefont{van~der Wiel,
  De~Franceschi, Elzerman, Tarucha, Kouwenhoven, Motohisa, Nakajima, and
  Fukui}}]{VanDerWiel02}
\bibinfo{author}{\bibfnamefont{W.~G.} \bibnamefont{van~der Wiel}},
  \bibinfo{author}{\bibfnamefont{S.}~\bibnamefont{De~Franceschi}},
  \bibinfo{author}{\bibfnamefont{J.~M.} \bibnamefont{Elzerman}},
  \bibinfo{author}{\bibfnamefont{S.}~\bibnamefont{Tarucha}},
  \bibinfo{author}{\bibfnamefont{L.~P.} \bibnamefont{Kouwenhoven}},
  \bibinfo{author}{\bibfnamefont{J.}~\bibnamefont{Motohisa}},
  \bibinfo{author}{\bibfnamefont{F.}~\bibnamefont{Nakajima}}, \bibnamefont{and}
  \bibinfo{author}{\bibfnamefont{T.}~\bibnamefont{Fukui}},
  \bibinfo{journal}{Phys. Rev. Lett.} \textbf{\bibinfo{volume}{88}},
  \bibinfo{pages}{126803} (\bibinfo{year}{2002}).

\bibitem[{\citenamefont{Nozi{\`e}res and Blandin}(1980)}]{Nozieres80}
\bibinfo{author}{\bibfnamefont{P.}~\bibnamefont{Nozi{\`e}res}}
  \bibnamefont{and} \bibinfo{author}{\bibfnamefont{A.}~\bibnamefont{Blandin}},
  \bibinfo{journal}{J. Phys. (Paris)} \textbf{\bibinfo{volume}{41}},
  \bibinfo{pages}{193} (\bibinfo{year}{1980}).

\bibitem[{\citenamefont{von Delft and Ralph}(2001)}]{VonDelft01}
\bibinfo{author}{\bibfnamefont{J.}~\bibnamefont{von Delft}} \bibnamefont{and}
  \bibinfo{author}{\bibfnamefont{D.}~\bibnamefont{Ralph}},
  \bibinfo{journal}{Phys. Rep} \textbf{\bibinfo{volume}{345}},
  \bibinfo{pages}{61} (\bibinfo{year}{2001}).

\bibitem[{\citenamefont{Beenakker}(1991)}]{Beenakker91}
\bibinfo{author}{\bibfnamefont{C.~W.~J.} \bibnamefont{Beenakker}},
  \bibinfo{journal}{Phys. Rev. B} \textbf{\bibinfo{volume}{44}},
  \bibinfo{pages}{1646} (\bibinfo{year}{1991}).

\bibitem[{fns()}]{fnspacing}
\bibinfo{note}{Assumption (1) and (2) do not affect the {\em spacing} between
  the peaks in $G$, but they do affect the relative peak heights at $B_{\rm
  Z}=0$. The peak spacings contain enough information to extract all the energy
  splittings illustrated in Fig.~\ref{fig:illust}.}

\bibitem[{EFt()}]{EFterm}
\bibinfo{note}{The energies $E_i[N]$ are assumed to contain a term originating
  from the equilibrium chemical potential, $E_F N$.}

\bibitem[{fnG()}]{fnGamma2Gamma1}
\bibinfo{note}{The case $\Gamma_2/\Gamma_1 = 1$ is clearly outside the region
  of validity of our $\Gamma_2/\Gamma_1 \gg 1$ treatment but is included in the
  plot nonetheless in order to make the trends clear.}

\bibitem[{\citenamefont{Aleiner et~al.}(2002)\citenamefont{Aleiner, Brouwer,
  and Glazman}}]{Aleiner02}
\bibinfo{author}{\bibfnamefont{I.~L.} \bibnamefont{Aleiner}},
  \bibinfo{author}{\bibfnamefont{P.~W.} \bibnamefont{Brouwer}},
  \bibnamefont{and} \bibinfo{author}{\bibfnamefont{L.~I.}
  \bibnamefont{Glazman}}, \bibinfo{journal}{Phys. Rep.}
  \textbf{\bibinfo{volume}{358}}, \bibinfo{pages}{309} (\bibinfo{year}{2002}),
  \bibinfo{note}{and references therein}.

\bibitem[{\citenamefont{De~Franceschi et~al.}(2001)\citenamefont{De~Franceschi,
  Sasaki, Elzerman, van~der Wiel, Tarucha, and
  Kouwenhoven}}]{DeFranceschiPRL01}
\bibinfo{author}{\bibfnamefont{S.}~\bibnamefont{De~Franceschi}},
  \bibinfo{author}{\bibfnamefont{S.}~\bibnamefont{Sasaki}},
  \bibinfo{author}{\bibfnamefont{J.~M.} \bibnamefont{Elzerman}},
  \bibinfo{author}{\bibfnamefont{W.~G.} \bibnamefont{van~der Wiel}},
  \bibinfo{author}{\bibfnamefont{S.}~\bibnamefont{Tarucha}}, \bibnamefont{and}
  \bibinfo{author}{\bibfnamefont{L.~P.} \bibnamefont{Kouwenhoven}},
  \bibinfo{journal}{Phys. Rev. Lett.} \textbf{\bibinfo{volume}{86}},
  \bibinfo{pages}{878} (\bibinfo{year}{2001}).

\bibitem[{\citenamefont{Zumb\"uhl et~al.}(2004)\citenamefont{Zumb\"uhl, Marcus,
  Hanson, and Gossard}}]{ZumbuhlPRL04}
\bibinfo{author}{\bibfnamefont{D.~M.} \bibnamefont{Zumb\"uhl}},
  \bibinfo{author}{\bibfnamefont{C.~M.} \bibnamefont{Marcus}},
  \bibinfo{author}{\bibfnamefont{M.~P.} \bibnamefont{Hanson}},
  \bibnamefont{and} \bibinfo{author}{\bibfnamefont{A.~C.}
  \bibnamefont{Gossard}}, \bibinfo{journal}{Phys. Rev. Lett.}
  \textbf{\bibinfo{volume}{93}}, \bibinfo{pages}{256801}
  (\bibinfo{year}{2004}).

\bibitem[{\citenamefont{Makarovski et~al.}(2006)\citenamefont{Makarovski, An,
  Liu, and Finkelstein}}]{Makarovski06}
\bibinfo{author}{\bibfnamefont{A.}~\bibnamefont{Makarovski}},
  \bibinfo{author}{\bibfnamefont{L.}~\bibnamefont{An}},
  \bibinfo{author}{\bibfnamefont{J.}~\bibnamefont{Liu}}, \bibnamefont{and}
  \bibinfo{author}{\bibfnamefont{G.}~\bibnamefont{Finkelstein}},
  \bibinfo{journal}{Phys. Rev. B} \textbf{\bibinfo{volume}{74}},
  \bibinfo{pages}{155431} (\bibinfo{year}{2006}).

\bibitem[{\citenamefont{Clogston and Anderson}(1961)}]{AndClog}
\bibinfo{author}{\bibfnamefont{A.~M.} \bibnamefont{Clogston}} \bibnamefont{and}
  \bibinfo{author}{\bibfnamefont{P.~W.} \bibnamefont{Anderson}},
  \bibinfo{journal}{Bull. Am. Phys. Soc.} \textbf{\bibinfo{volume}{6}},
  \bibinfo{pages}{124} (\bibinfo{year}{1961}).

\bibitem[{\citenamefont{Barzykin and Affleck}(1998)}]{Barzykin98}
\bibinfo{author}{\bibfnamefont{V.}~\bibnamefont{Barzykin}} \bibnamefont{and}
  \bibinfo{author}{\bibfnamefont{I.}~\bibnamefont{Affleck}},
  \bibinfo{journal}{Phys. Rev. B} \textbf{\bibinfo{volume}{57}},
  \bibinfo{pages}{432} (\bibinfo{year}{1998}).

\bibitem[{Fin()}]{FiniteSize_chiloc}
\bibinfo{note}{In a finite system, a naive estimate gives a $\Delta_R/T_K$
  finite size correction to the local susceptibility from the second term in
  Eq.\,(\ref{eq:Epsi_strong}).}

\bibitem[{can()}]{canon_method}
\bibinfo{note}{The spin-chain mapping \cite{WilsonRMP75,HewsonBook} of
  Ref.\,\onlinecite{Yoo05} is easily modified to simulate the canonical
  ensemble for the Kondo problem. Fixing the number of fermions in the Kondo
  box corresponds to simulating the spin-chain at fixed magnetization, while
  the grand-canonical case corresponds to a fixed magnetic field. During the
  loop update, the magnetization of the spin-chain is updated when loops wrap
  around the imaginary time direction (they have temporal winding). If we
  prevent our loops from winding temporally, the magnetization of the
  spin-chain stays fixed, and we are simulating the canonical ensemble in the
  Kondo problem. A simple approach to restricting loops to the zero temporal
  winding sector is to finish the growth of a loop and then reject loops that
  wind in time. A more efficient approach and the one that we have used here is
  to force the loops to bounce back if they cross a certain time slice. It is
  possible to show that this algorithm satisfies detailed balance and at the
  same time forbids temporal winding.}

\end{thebibliography}

\end{document}